\documentclass[a4paper,usenatbib]{mnras}

\usepackage{newtxtext,newtxmath}
\usepackage{longtable}
\usepackage{lscape} 
\usepackage{siunitx}

\usepackage[T1]{fontenc}
\usepackage{ae,aecompl}

\usepackage{graphicx}	
\usepackage{amsmath}	
\usepackage{amssymb}	
\usepackage{caption}
\title[CMEs on solar like stars]{A census of Coronal Mass Ejections on solar-like stars \thanks{Based on data obtained from the ESO Science Archive Facility under request number mal394399, mal394575, mal394457, mal394506, mal394559, mal394761, mal394843, mal394848, mal394945, mal395064}}

\author[M. Leitzinger et al.]{
M. Leitzinger,$^{1}$\thanks{E-mail: martin.leitzinger@uni-graz.at}
P. Odert,$^{1}$
R. Greimel,$^{1}$
K. Vida,$^{2}$
L. Kriskovics,$^{2}$
E.W. Guenther,$^{3}$
\newauthor
H. Korhonen,$^{4}$
F. Koller,$^{1}$
A. Hanslmeier,$^{1}$
Zs. K\H{o}v\'ari,$^{2}$
H. Lammer,$^{5}$
\\
$^{1}$Institute of Physics/IGAM, University of Graz, Universit\"atsplatz 5, A-8010 Graz, Austria\\
$^{2}$Konkoly Observatory, MTA CSFK, H-1121 Budapest, Konkoly Thege M. \'ut 15-17, Hungary\\
$^{3}$Th\"uringer Landessternwarte Tautenburg, Sternwarte 5, D-07778 Tautenburg, Germany \\
$^{4}$Dark Cosmology Centre, Niels Bohr Institute, University of Copenhagen, Juliane Maries Vej 30, DK-2100 Copenhagen, Denmark\\
$^{5}$Space Research Institute, Austrian Academy of Sciences, Schmiedlstra\ss{}e 6, A-8042, Graz, Austria
}
\date{Accepted XXX. Received YYY; in original form ZZZ}

\pubyear{2019}

\begin{document}
\label{firstpage}
\pagerange{\pageref{firstpage}--\pageref{lastpage}}
\maketitle

\begin{abstract}
Coronal Mass Ejections (CMEs) may have major importance for planetary and stellar evolution. Stellar CME parameters, such as mass and velocity, have yet not been determined statistically. So far only a handful of stellar CMEs has been detected mainly on dMe stars using spectroscopic observations. We therefore aim for a statistical determination of CMEs of solar-like stars by using spectroscopic data from the ESO phase 3 and Polarbase archives. To identify stellar CMEs we use the Doppler signal in optical spectral lines being a signature of erupting filaments which are closely correlated to CMEs. 
We investigate more than 3700~hours of on-source time of in total 425 dF-dK stars. 
We find no signatures of CMEs and only few flares. To explain this low level of activity we derive upper limits for the non detections of CMEs and compare those with empirically modelled CME rates. To explain the low number of detected flares we adapt a flare power law derived from EUV data to the H$\alpha$ regime, yielding more realistic results for H$\alpha$ observations. In addition we examine the detectability of flares from the stars by extracting Sun-as-a-star H$\alpha$ light curves. 
The extrapolated maximum numbers of observable CMEs are below the observationally determined upper limits, which indicates that the on-source times were mostly too short to detect stellar CMEs in H$\alpha$. We conclude that these non detections are related to observational biases in conjunction with a low level of activity of the investigated dF-dK stars.

\end{abstract}




\section{Introduction}
Stellar activity is driven by magnetic energy and has various forms. Two of the most energetic activity phenomena are sudden outbreaks of radiation detectable more or less all over the electromagnetic spectrum (flares) and coronal mass ejections (CMEs) which are expulsions of highly energetic particles such as fast electrons and protons. Both phenomena are very well studied on the Sun especially when compared to stars. For solar observations massive effort has been conducted to monitor the Sun since decades. This enables the determination of parameters of these solar activity phenomena. This allows the determination of mass, velocity, and kinetic energy of solar CMEs and their correlation with associated phenomena such as flares and coronal dimmings (evacuated regions in the corona after a CME was ejected) with unprecedented statistical significance \citep[e.g.][and references therein]{Vourlidas2010, Aarnio2011, Youssef2013, Compagnino2017, Dissauer2019}. Moreover, observations of solar activity phenomena allow of course also investigations with high spatial resolution. On stars, flaring regions or CMEs can not be spatially resolved and stars are not monitored over decades as it is the case for the Sun. Although there exist long-term studies \citep[e.g.][]{Wilson1968, Baliunas1995, Gray2015} but those are done with low frequency. \\
When stars enter the main-sequence (MS) their level of activity, i.e. spottedness, coronal (X-ray, EUV)- and chromospheric (e.g. Balmer lines) emission,  is high \citep[e.g.][]{Guedel2007}, and several of these indicators exhibit a saturation effect \citep[see e.g.][]{Pallavicini1981, Pizzolato2003, Wright2011, Jackson2012, Reiners2014, Tu2015, Wright2018}. As stars evolve on the MS the level of activity decreases due to stellar spin-down which is caused by magnetic braking because of a magnetized wind \citep[e.g.][]{WeberDavis1967, Skumanich1972, Kawaler1988, Matt2012, Brown2014, Garraffo2018, See2018}. This is evident from investigations of the high-energy radiation of solar analogue stars of different age (see e.g. \citet{Ribas2005} for solar analogue stars. Accordingly also the flare frequency  decreases \citep[e.g.][]{Audard2000}. Even in the optical range this has been found \citep[see e.g.][]{Davenport2016, Balona2015}.\\
Regarding mass outflow processes the stellar picture is more ambiguous. 
Radio observations of solar analogue stars of different age have set upper limits to the mass loss rates of those stars \citep{Lim1996, Gaidos2000, Fichtinger2017} due to the non-detection of free-free emission as a signature of a hot ionized wind from those stars. Another approach to measure stellar mass loss of MS stars has been presented by \citet{Linsky1996, Wood2004} who use the interaction of the stellar wind with the interstellar medium as a method (astrospheric absorption) to determine stellar mass loss. Those authors find a relation between stellar X-ray flux and mass loss. This relation holds for stars with ages down to 0.5~Gyr. Younger stars do not follow this relation. For instance, the 0.3~Gyr old solar analogue $\pi^{1}$~UMa shows a mass loss rate even lower than that of the present-day Sun \citep{Wood2014, Wood2018}. This is even more surprising as this star has a much more active chromosphere than the present day Sun. 
The weak wind of $\pi^{1}$~UMa as determined from astrospheric absorption does not contradict the upper limits derived from radio observations \citep{Fichtinger2017} as the radio approach gives an upper limit of the mass loss rate in the order of 10$^{-12}$M$_{Sun}$~year$^{-1}$ whereas the astrosphere method gives a mass loss rate in the order of 10$^{-14}$M$_{Sun}$~year$^{-1}$.\\
The \textbf\textit{{second mass-loss driving mechanism}} - stellar CMEs - remains ambiguous as well. So far only a handful of detections in the optical using the method of Doppler shifted emission/absorption around Balmer lines is known from literature where the deduced projected velocity exceeded the escape velocity of the stars, i.e. stellar material was indeed ejected into the astrosphere \citep{Houdebine1990, Gunn1994, Guenther1997, Vida2016}. The method of Doppler shifted emission/absorption in Balmer lines is the signature of eruptive filaments which are closely correlated to CMEs on the Sun \citep{Hori2002,Gopalswamy2003}. \citet{FuhrmeisterSchmitt2004, Vida2019} also found optical signatures of possible stellar CMEs, but the deduced projected velocities did not exceed the escape velocities of the stars, Also from shorter wavelengths CME detections have been claimed. \citet{Bond2001} found in Hubble UV spectra of the pre-cataclysmic binary V471~Tau repeatedly absorption transients which they attributed to CMEs from the dK star component of the system. Moreover, this is the only star so far with a deduced CME rate (100-500~CMEs day$^{-1}$), which makes it an interesting target for the search for optical signatures of stellar CMEs. \citet{Leitzinger2011a, Argiroffi2019} found Doppler-shifted emission features at FUV and X-ray wavelengths, respectively, both being well below the escape velocities of the investigated stars. Several dedicated search programs for stellar CMEs in the optical did not show convincing results \citep{Leitzinger2014, Korhonen2017, Fuhrmeister2018}. \\
Beside the method of Doppler shifted emission/absorption of spectral lines in optical, FUV, and X-ray domains (as described above) there were also several attempts to interpret detected continuous X-ray absorptions as CMEs temporarily obscuring the star \citep{Haisch1983, OttmannSchmitt1996, Tsuboi1998, FavataSchmitt1999, Franciosini2001, Covino2001, Pandey2012}. A prominent representative and one of the most energetic X-ray flares to date was detected on Algol \citep{FavataSchmitt1999}. During this flare a sharp rise in column density was detected together with a continuous X-ray absorption decay which is interpreted as an expanding CME. \citet{Moschou2017} derived properties of the potential CME and put it in the context of the solar CME mass/flare energy relation, where the event fits very well. Moreover, \citet{Moschou2019} use the so far detected stellar CMEs found by both methods (Doppler shifted emission/absorption and X-ray absorptions) to investigate the stellar flare-CME relation, and derive, using a simple CME model, mass and energy for the CME events from literature. They find that stellar CMEs are more limited in velocity than in mass which is interpreted by the authors as a possible consequence of strong overlying magnetic fields.\\
Finally, also in the radio domain there were several attempts to search for stellar CMEs using type II bursts as a signature of a shock wave driven by CMEs, but no definite detection of stellar CMEs using this method has been reported so far \citep[see e.g.][and references therein]{Leitzinger2009, Boiko2012, Osten2015, Villadsen2017, Crosley2018a, Crosley2018b}. Only recently, \citet{Mullan2019} suggested that on active stars, CMEs are not able to generate shock-waves, thus no type II burst are triggered, making them radio-quiet. This would explain the non-detections reported so far.\\ 
On the Sun we know that CMEs make $\sim$10\% of the total solar mass loss per year \citep{Lamy2017, Mishra2019}. So the question:``Could the wind of young MS stars be dominated by CMEs?'' can be posed. Both above mentioned wind detection methods (radio and astrospheric absorption) can not distinguish between wind and a superposition of numerous CMEs. So the only way to answer this question is to determine stellar CME parameters (mass, frequency) on a statistical basis. But not only stellar mass loss is affected by CMEs, also stellar angular-momentum loss is closely connected to CMEs, probably playing an important role in stellar spin-down. As known from the Sun there is interaction between CMEs and the Earth's atmosphere. Model results for exoplanet systems have shown that frequent and energetic CMEs together with a high short-wavelength (EUV) radiation environment lead to very efficient planetary atmospheric mass loss \citep{Lammer2007, Khodachenko2007, Cohen2011, Cherenkov2017, Johnstone2019}. In a worst-case scenario frequent and energetic CME impacts may lead to the loss of planetary atmospheres and therefore CMEs endanger planetary habitability.

\section{Observations}
\subsection{Data sources}
We select the Phase 3 archive\footnote{\url{http://archive.eso.org/wdb/wdb/adp/phase3_main/form}} of the European Southern Observatory (ESO) and the Polarbase archive\footnote{\url{http://polarbase.irap.omp.eu/}} \citep{Petit2014} to search for optical signatures of CMEs on F-K main-sequence stars. It turned out that the High Accuracy Radial velocity Planet Searcher (HARPS) mounted on the 3.6~m telescope in LaSilla yielded the largest number of target stars (see section~\ref{targetstars}), compared to other ESO instruments in the phase 3 archive. The spectral resolving power of HARPS is 115000.  
  \begin{figure}
   \centering
    \includegraphics[width=8cm]{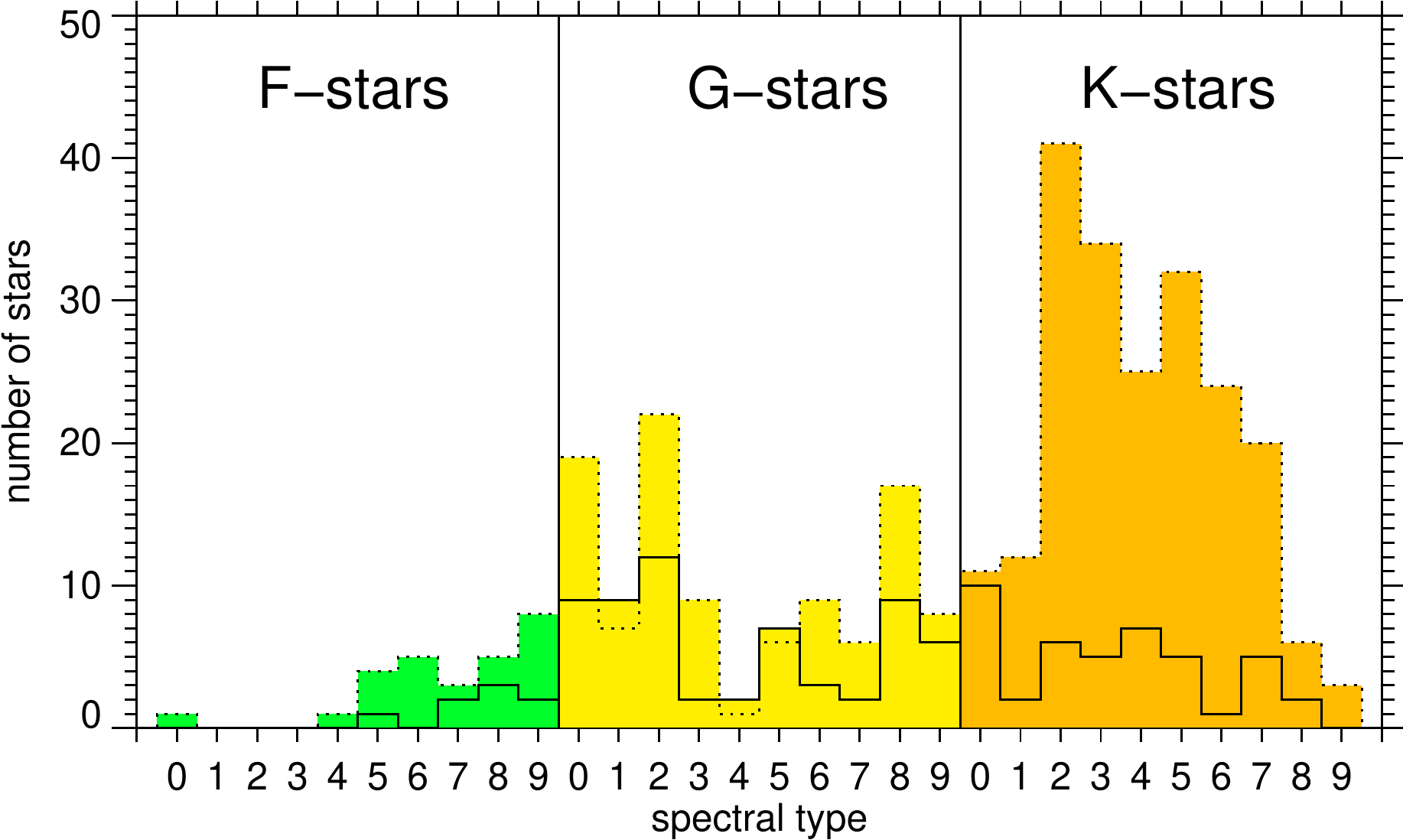} 
    \includegraphics[width=8cm]{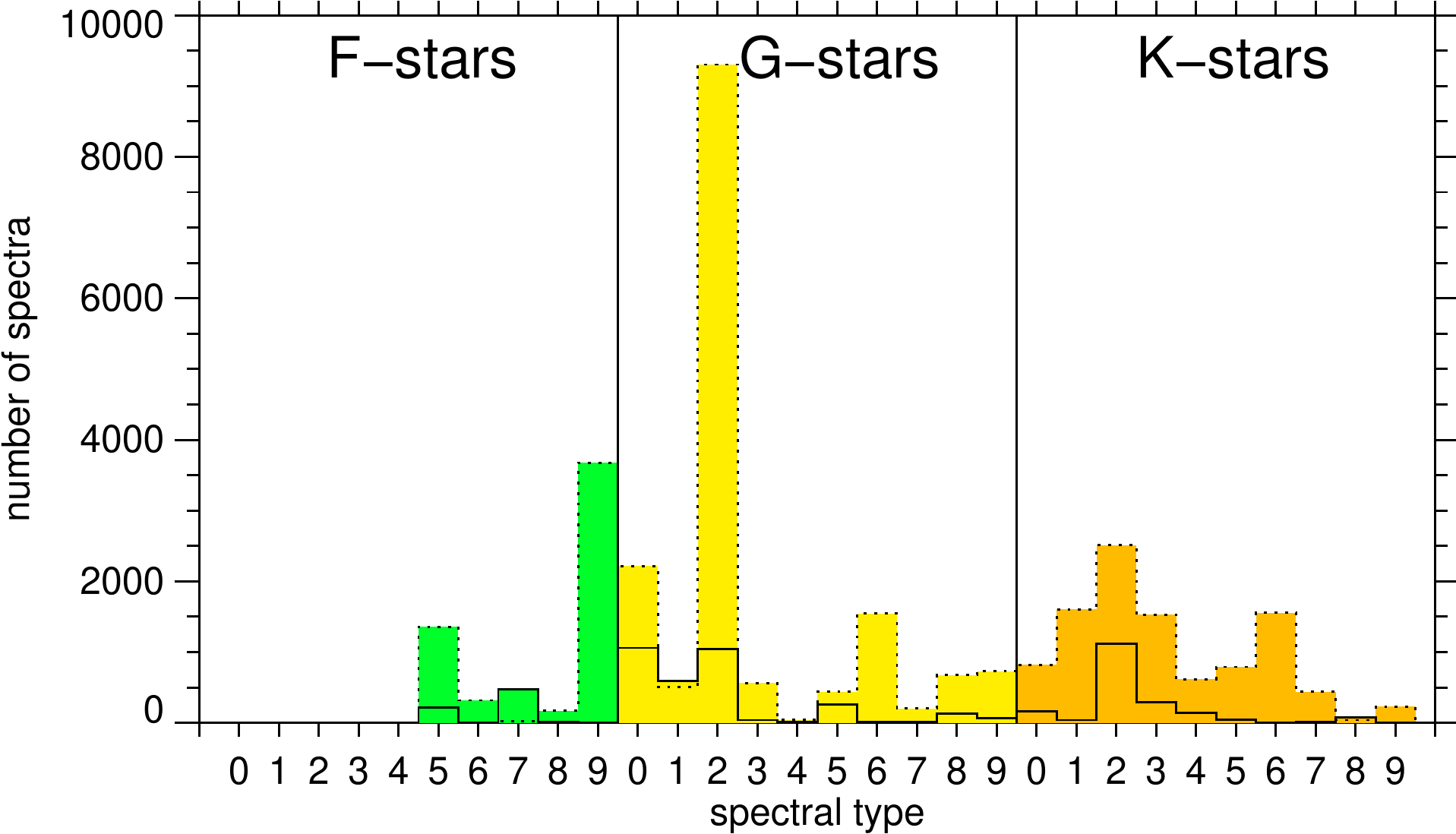}
  
   

     \caption{Upper panel: Histogram of the target stars. The solid line represents the distribution of the Polarbase targets whereas the dotted line represents the distribution of the HARPS targets. The most numerous stars in the sample are dK stars (251 from which 12 stars are redundant in both, ESO and Polarbase, archives), followed by dG stars (171 from which 15 are redundant in both, ESO and Polarbase, archives), and dF stars (34 from which 4 are redundant in both, ESO and Polarbase, archives). Lower panel: Histogram of the target spectra.
              }
         \label{Fig1}
   \end{figure}
The Polarbase archive includes spectral observations from the Echelle Spectro Polarimetric Device for the Observation of Stars (ESPaDOnS) instrument installed on the 3.6~m Canada-France-Hawaii Telescope (CFHT) and the Narval instrument installed on the 2.0~m Telescope Bernard Lyot. The spectral resolving power of ESPaDOnS ranges from 68000-81000 depending on polarimetric and object mode. The spectral resolving power of Narval ranges from 65000-75000 depending on different observing modes. The majority of the spectra were recorded as a continuous 4-spectra time series over a period of several years (see lower panel of Fig.~\ref{Fig1}). The datasets for each star are therefore no continuous time series. The majority of the HARPS spectra were recorded as single spectra. \\
\subsection{The target sample}
\label{targetstars}
\begin{figure}
  \centering
  \includegraphics[width=8cm]{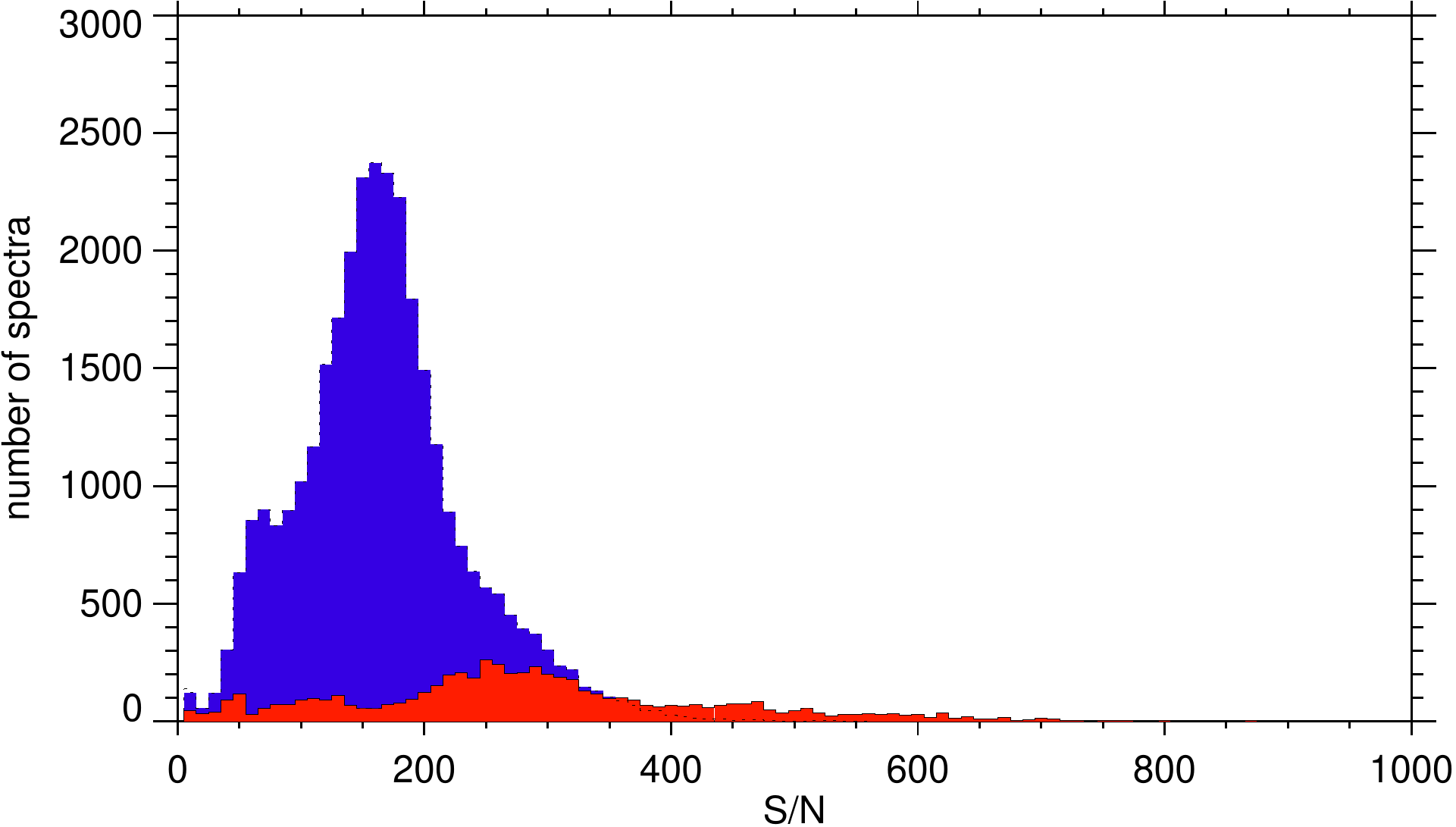}
  \includegraphics[width=8cm]{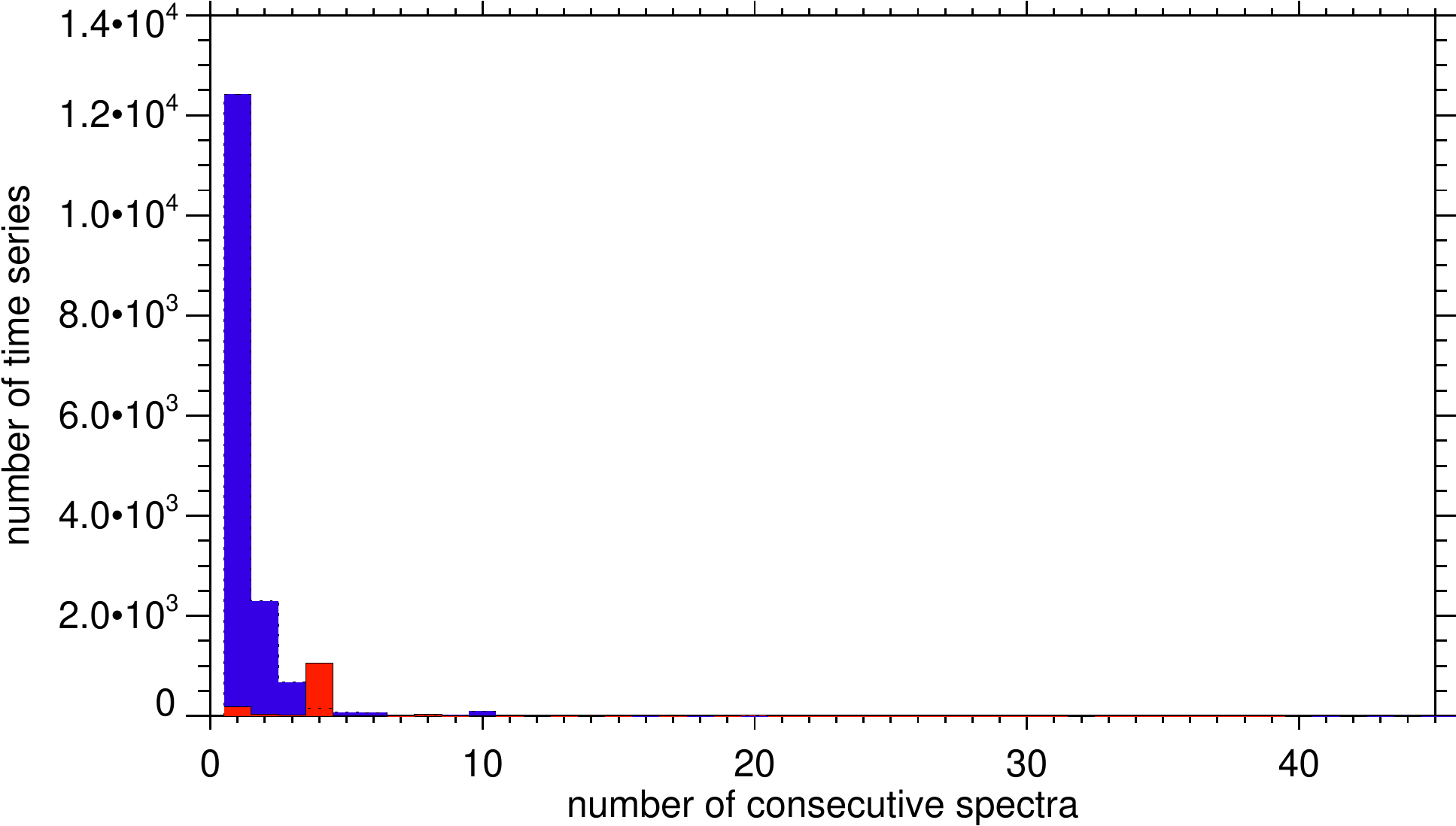} 
    
  \caption{Upper panel: Histogram of the S/N of the Polarbase (red colored distribution and solid line) and HARPS (blue colored distribution and dotted line) spectra.  
  Lower panel: Histogram of the number of consecutive spectra building a sub-time-series. Colors and lines have the same assignment as in the upper panel.}
         \label{Fig2}
   \end{figure}

In the present study we focus on solar-like stars (spectral types F-K). As a base for target star selection we use the catalogue of \citet{Hinkel2017} who compiled a catalogue of 951 F-K stars within 30~pc. Although the catalogue is not complete, as it is based on the Tycho-Gaia Astrometric Solution which has an 80\% completeness with respect to Hipparcos and Tycho-2, it represents a solid base of target stars of spectral type F-K in the solar neighbourhood. Because the catalogue is not complete, we add the target sample from the ``Sun in time'' project \citep[see e.g.][]{Guedel2007} as well as the target sample of \citet{Gaidos1998} to include the prominent young and active stars from those catalogs.\\
We do not restrict the target sample to single stars. In Table~\ref{appendixtable} we show the final target sample of the present study which have spectra in the Polarbase archives. From the initial target sample of 986 F-K stars we found 112 stars in the Polarbase and 344 stars in the ESO Phase 3 HARPS archive each with a different number of spectra (see~Table~\ref{appendixtable}, and Fig.~\ref{Fig1}). In total 31 stars appear in both, the Polarbase and the HARPS archive. The distribution shown in Fig.~\ref{Fig1} includes stars appearing in both archives. The total number of spectra of target stars from both archives gives 43229 spectra (Polarbase: 5468 spectra, HARPS: 37761 spectra). This translates to a total on-source time of $\sim$ 70 hours in the case of Polarbase and $\sim$ 3660 hours in the case of HARPS.  In Fig.~\ref{Fig2} we show the histograms of the observing parameters. In the upper panel of Fig.~\ref{Fig2} we show the distribution of the different signal-to-noise (S/N) values of the spectra (blue colored distribution: HARPS, red colored distribution: Polarbase). The S/N values of the Polarbase spectra vary between 10 and 984 with a mean of 296, while in case of the HARPS spectra, it ranges from 10 to 558, yielding a mean of 161.
In the lower panel of Fig.~\ref{Fig2} we show the distribution of time series consisting of consecutive spectra (sub-time-series). The majority of Polarbase spectra (red colored distribution) were recorded as 4-spectra sub-time-series whereas the majority of HARPS spectra (blue colored distribution) were recorded as single spectra.\\
 
\subsection{Data preparation}
The data in the Polarbase archive are offered as reduced normalized and not-normalized spectra merged from all Echelle orders of the respective spectrographs. The data in the ESO phase 3 HARPS archive are also offered as reduced data. We use the HARPS advanced data products (adp) offered by ESO. The single Echelle orders are merged into one data file per exposure. To search for extra emissions/absorptions related to CMEs we use the common method of comparing individual spectra to a quiescent spectrum constructed by averaging the spectra of each times series \citep[see e.g.][]{Leitzinger2011a, Leitzinger2014, Vida2016, Vida2019}. For this comparison the spectra need to be continuum normalized because every spectrum needs to have the same continuum level to be compared to other spectra and the quiescent spectrum. In the following we describe how we normalized the spectra.\\ 
\textbf{\textit{Continuum normalization:}} We determine the continuum of a star by examining its quiescent spectrum. To build a quiescent spectrum from a spectral time series we fit the individual spectra with a straight line excluding strong spectral lines (e.g. Balmerlines) which would affect the continuum fit. To avoid contamination by very noisy spectra (i.e. spectra with S/N$<$15) we discard those from the process of building the quiescent spectrum. The quiescent spectrum is then generated by calculating the arithmetic mean of the remaining spectra of the timeseries. From this quiescent spectrum we determine the continuum which is then used to normalize each spectrum of the time series. By selecting this approach we can be sure that any deviation of a spectrum from the quiescent spectrum is not related to continuum normalization. If we normalized every spectrum by its individual continuum we remove possible broad emissions/absorptions related to CMEs. For continuum determination of the quiescent spectrum we tested polynomial fitting of spectra including local maxima of spectra and local maxima of spectra only \citep[cf.][]{Zhao2006}. Polynomial fitting of various degrees of the spectra and the local maxima of the spectra, resulted in a wave-like normalized continuum. Division of the quiescent spectrum by its local maxima without polynomial fitting yielded a flat continuum. Therefore we adopted this method. Every spectrum of the time series of each star was then divided by its continuum, determined from the local maxima, from the quiescent spectrum of the corrsponding star. The reason why we aim for a flat continuum is, that when overplotting the times series of spectra and the quiescent spectrum, deviations from the quiescent spectrum are easier visible if the continuum is flat. To summarize the normalization procedure, we build for every star a quiescent spectrum, determine its continuum and divide each spectrum of the corresponding spectral time series by this continuum.  We then generate plots which show the quiescent and the individual spectra. By visual inspection we are then able to identify signatures of CMEs. To detect signatures of CMEs the CME must produce a sufficient amount of flux (larger than the noise in the spectra) and a width $>$1\AA{}, which is roughly width of stellar quiescent prominences \citep{CollierCameron1989a}. Every so far detected stellar CME in optical spectra had a width in wavelength of $>$5\AA{} \citep[cf.][and references therein]{Moschou2019}, which can be explained by the expansion of a CME when it is ejected \citep[see e.g.][and references therein]{Howard2015, Vida2019}.\\
\textbf{\textit{Errors:}} The Polarbase data include error estimations. The HARPS ESO Phase 3 data products do not include error estimations, therefore we determine the S/N from the raw 1D spectra following the approach given in \citet{Dumusque2018}. Here the noise in HARPS spectra is determined from the photon noise (square root of the flux) and the CCD read-out and dark current noise which is 12 photo-electrons for HARPS. The flux of each adp spectrum is then divided by its S/N, which gives then the error of each spectrum. For data preparation and analysis steps the errors are propagated using the standard technique of Gaussian error propagation. For the data preparation as well as for the analysis we use the Interactive Data Language (IDL).\\
As the HARPS adp files are already order merged we additionally use the HARPS e2ds files for sanity checks of the corresponding adp files. The HARPS e2ds files are extracted two dimensional spectra. The files contain the single Echelle orders in photo-electrons unit. In this way we are able to identify any merging problem probably having occurred during the generation process of the HARPS adp files.

\section{Results}
\subsection{CMEs}
\label{CMEres}
\begin{figure*}
  \centering
   \includegraphics[width=17cm]{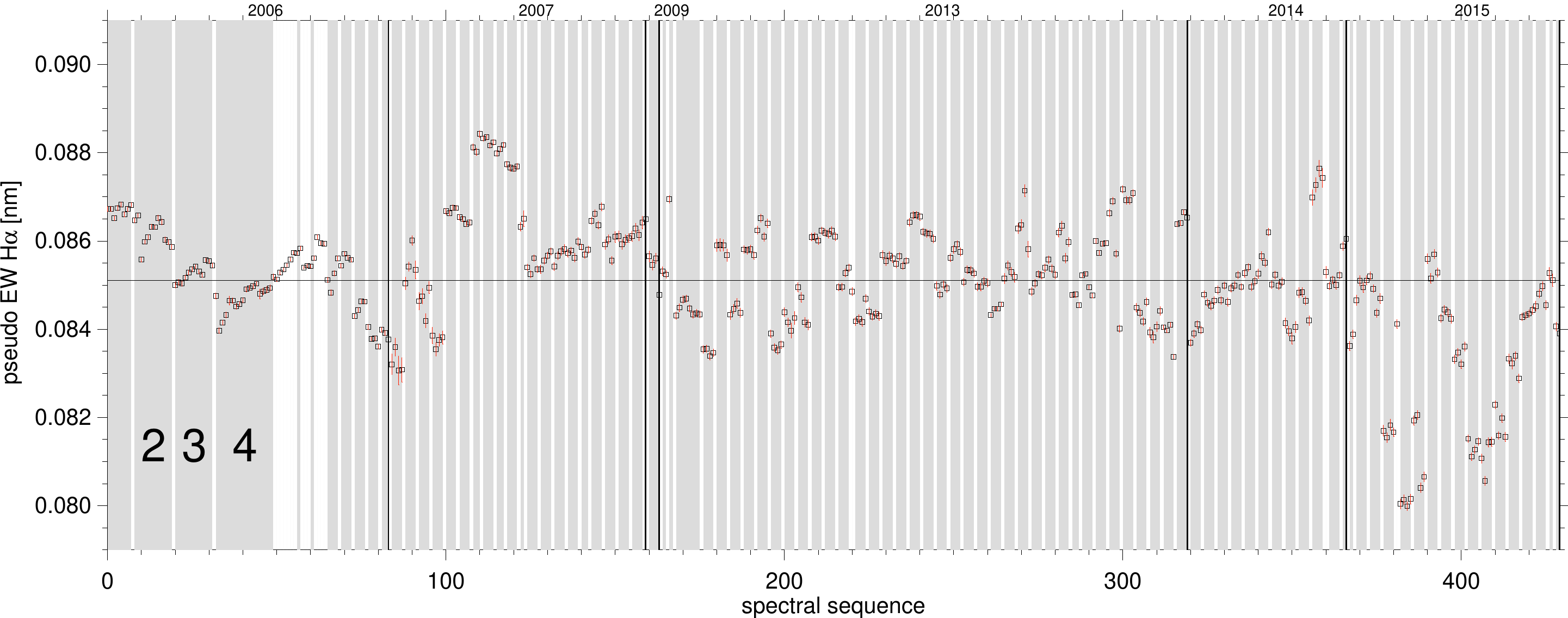} 
   \caption{Spectral sequence versus pseudo EW of HD189733 (TYC2141-972-1). The grey shaded areas denote the sub-time-series. The vertical solid black lines denote the dividing lines between years of observation. The horizontal solid black line denotes the average of the EWs shown.}
         \label{Fig3}
   \end{figure*} 
%
%
   
For the search for signatures of stellar CMEs we focus on two Balmer lines, namely H$\alpha$ and H$\beta$.  Hereby we assume that erupting filaments are visible in H$\alpha$ and H$\beta$. 
From the visual inspection we found no signatures of stellar CMEs. For the stars which show H$\alpha$ in emission (in total 9 stars, see Table~\ref{appendixtable}), which should be more active than the stars which show H$\alpha$ in absorption, we find only marginal wing asymmetries lying within the noise of the corresponding spectra. \\
We also analyzed the HARPS e2ds files and found also no signature of stellar CMEs, in agreement with the HARPS adp analysis.\\
As we did not find any signatures of stellar CMEs we estimate upper limits of the CME rates. 
To estimate the upper limits we use the relation -ln(1-0.95)/t$_{\mathrm{obs}}$ \citep{Gehrels1986} which gives the 95\% confidence limit. The upper limits of the CME rates are given in Table~\ref{appendixtable}.\\
\subsection{Flares and chromospheric activity}
\label{flaresres}
\begin{figure}
  \centering
  \includegraphics[width=\columnwidth]{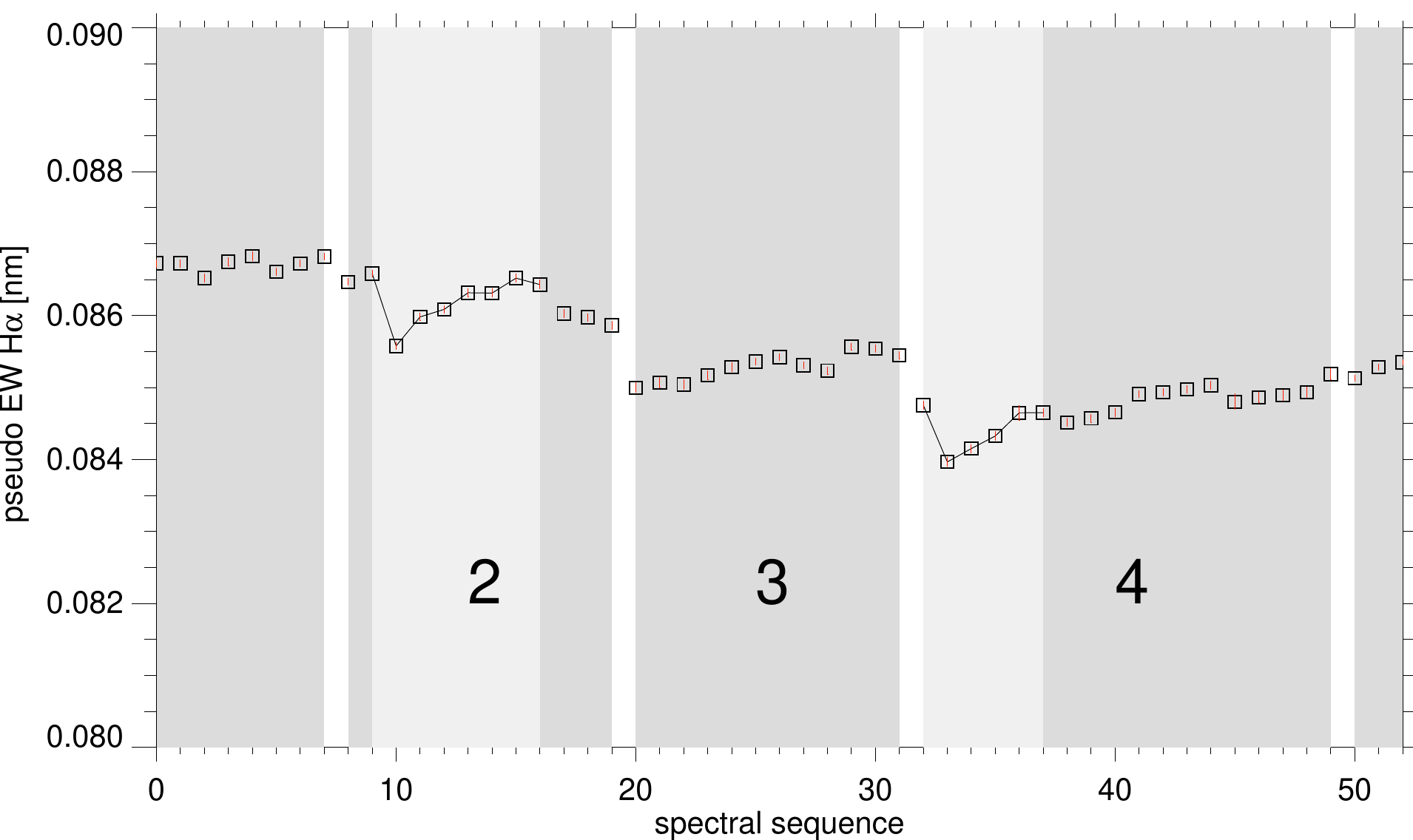}
   \caption{Cut-out of the spectral sequence of HD189733. In sub-time-series 2 and 4 the pseudo EWs show a typical flaring behaviour (light-gray shading).}
         \label{Fig4}
   \end{figure}

\begin{figure}
  \centering
   \includegraphics[width=\columnwidth]{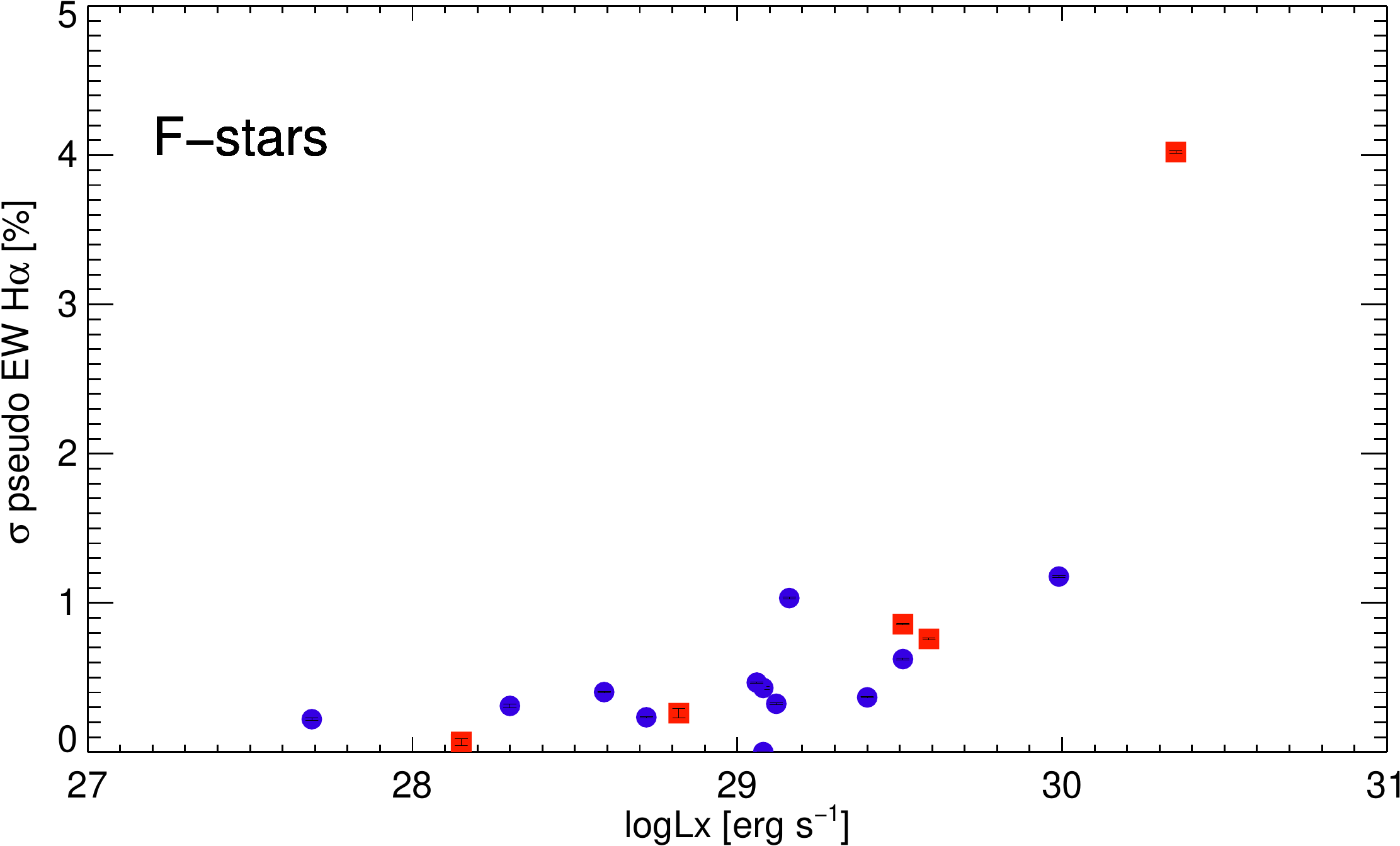}
   \includegraphics[width=\columnwidth]{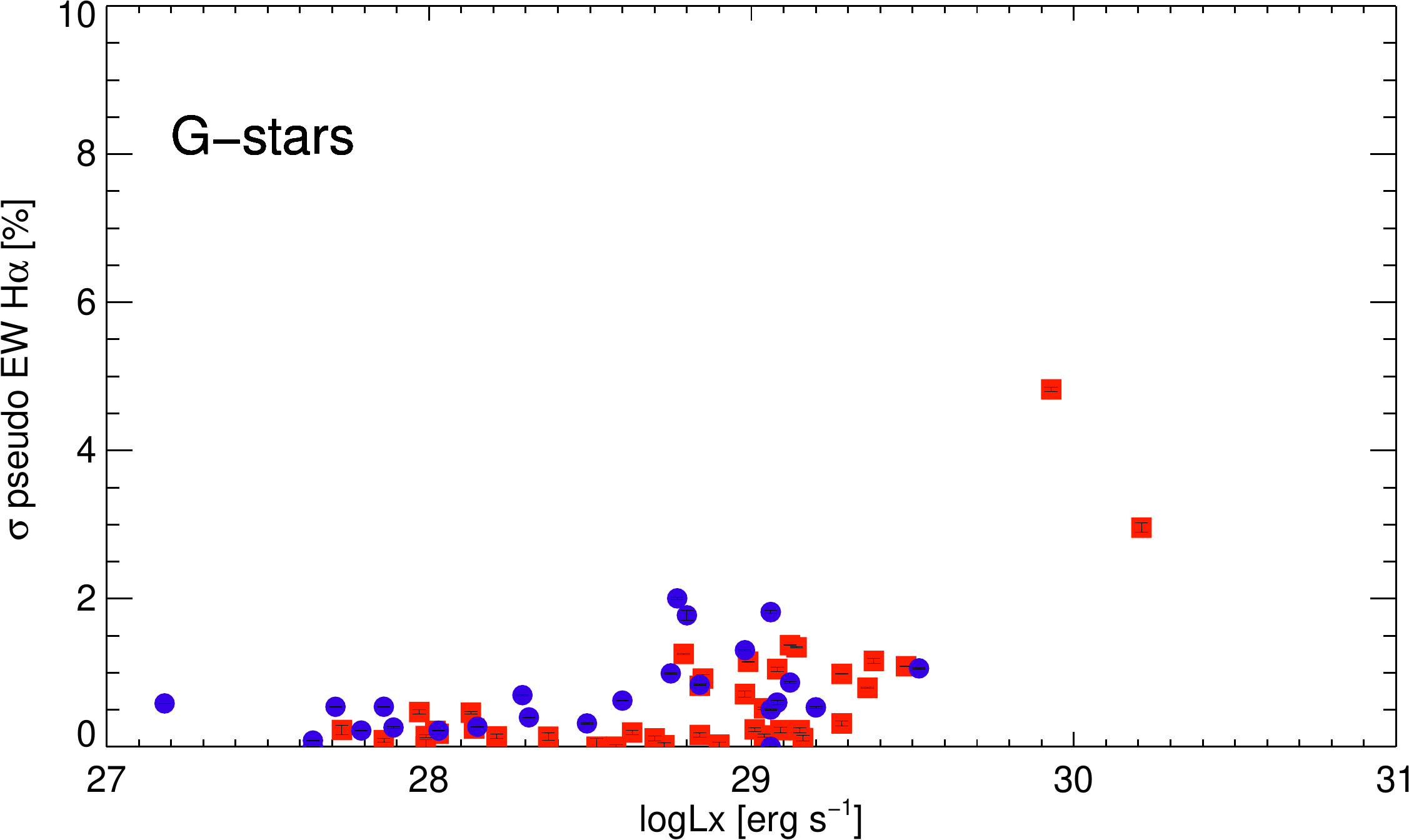}
   \includegraphics[width=\columnwidth]{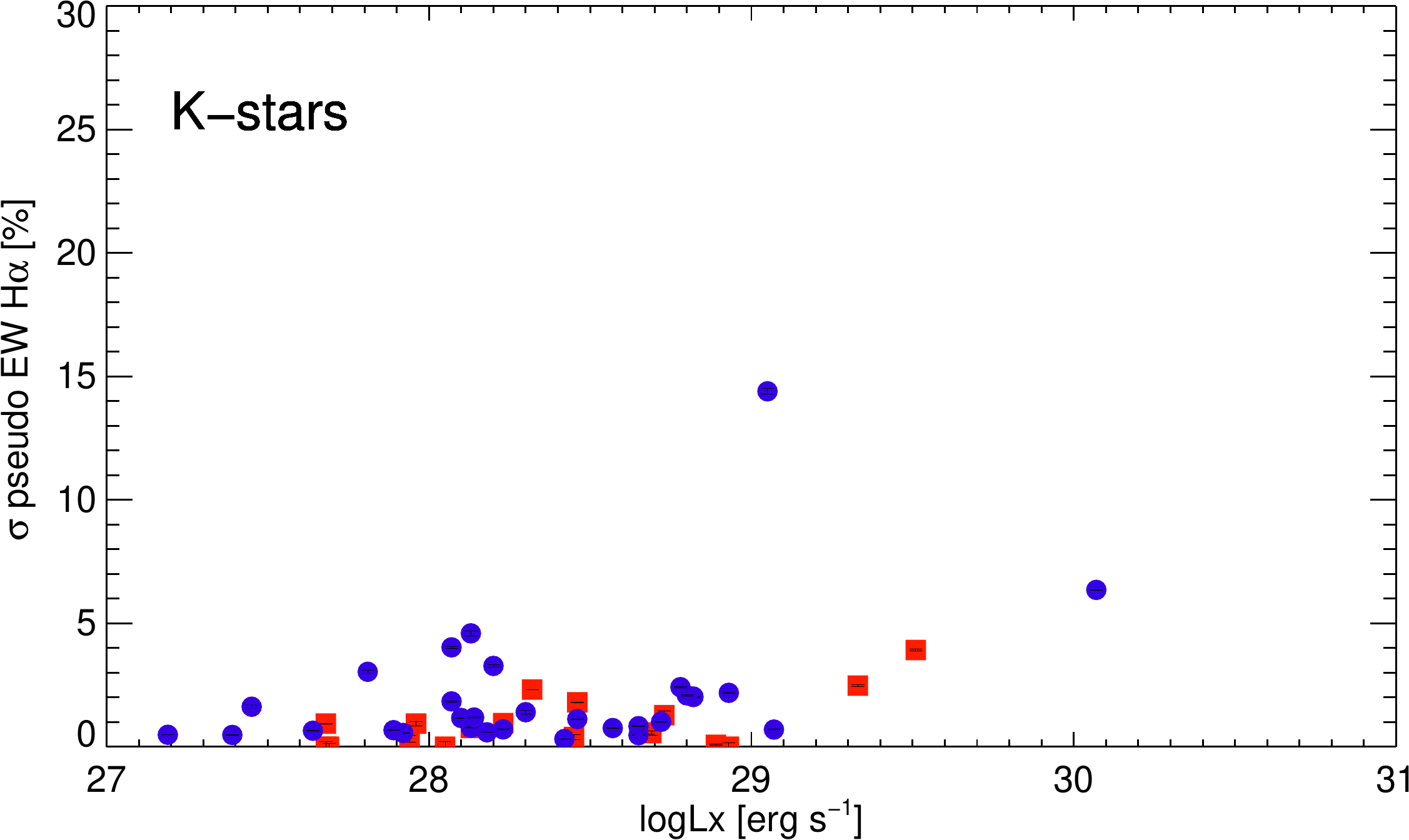}
   \caption{$\log L_{\mathrm{X}}$ versus $\sigma$ of pseudo H$\alpha$ EW in \% of the corresponding pseudo EW 
   mean values of all F-, G, and K-stars of the sample with more than 20 spectra in the Polarbase and HARPS archive. Blue dots denote sigma of pseudo EWs extracted from HARPS data and red squares from Polarbase data. }
         \label{Fig5}
   \end{figure}

The stars of the sample showing H$\alpha$ in absorption do not show variations of the wings (see section~\ref{CMEres}). But we find variations of the line cores of the chromospheric lines in a number of stars. As the majority of the spectral time series in the Polarbase archive is 4 spectra long, and in the HARPS archive even smaller (see~Fig.~\ref{Fig2}), it is difficult to determine whether the variations of the H$\alpha$ cores originate from flares or hemispheres with a different population of active regions facing the observer. This becomes even more obvious when we plot pseudo EWs of the timeseries of each star. We define the pseudo EW by a 2\AA{} wide wavelength window centered on the H$\alpha$ line core. The HARPS spectra are recorded simultaneously with a wavelength calibration lamp necessary for the determination of accurate radial velocity, and furthermore, the spectra, both from the Polarbase and HARPS archive, are contaminated by telluric lines. To avoid the correction for both the spectral lines of the HARPS wavelength calibration lamp and telluric lines, we decided to use a fixed and narrow wavelength window, for which the determined pseudo EW does not need to be corrected for those.\\
As an example we show the variation of pseudo EWs of the 600~Myr old solar-like star HD189733. As one can see the variation of the H$\alpha$ pseudo EW over years (the timeseries spans the range from 2006 until 2015) accounts to about $\pm$5\% of the mean pseudo EW as indicated with a horizontal solid black line in Fig.~\ref{Fig3}. In all young stars in our sample we see these variations in pseudo EW.\\
As noted above, the identification of flares is mainly hampered by the fact that the majority of the duration of the spectral sub-time-series is four spectra or smaller. If one takes a closer look at Fig.~\ref{Fig3} one recognizes that in this data set there are seven sub-time-series which have a length of more than the typical four spectra. In the second and fourth of those, occurring from spectral sequence 8-49, we identify typical flaring behaviour of a star showing H$\alpha$ in absorption, with a typical rapid decrease and slower increase in EW (see Fig.~\ref{Fig4}). For all the remaining stars from the Polarbase archive analysed in the present study we did not detect any flares. The stars in the HARPS archive analysed in the present study show one flare on the star $\iota$~Hor.\\
To calculate the flare energies of the flares on those stars we use continuum fluxes of $\epsilon$~Eri for HD~189733 and HD~45067 for $\iota$~Hor, which have the same spectral type, from the stellar spectral library from \citet{Cincunegui2004}. The continuum fluxes of \citet{Cincunegui2004} are then multiplied by the corresponding normalized spectra and also accounting for the difference in distance of $\epsilon$~Eri and HD~1897333 and HD~45067 and $\iota$~Hor. The flare energy is then calculated by integrating over the pseudo EW according to E = 4 $\pi$ d$^{2}$ f$_{cont}$ $\int$ (EW-EW(t=0))dt. This yields then lower limit energies (because we do not consider the whole spectral H$\alpha$ profile for the determination of the EW) of $\sim$3$\times$10$^{27}$erg for Flare 1 and $\sim$2$\times$10$^{27}$erg for Flare 2 on HD~189733 and $\sim$1.1$\times$10$^{29}$erg for the flare on $\iota$~Hor.\\
We have found only 3 flares on two out of 425 stars when we identify flares from their evolution with a short rise and a longer decay phase. This of course works only for sub-time-series which consist of more than 1 spectrum. If we define a flare by its peak being significantly higher than the quiescent light curve (mean of the pseudo EWs) of a star then the flare search in sub-time-series with one spectrum becomes feasible \citep[cf.][]{Pavlenko2019}. Defining a significant enhancement relative to the quiescent light curve as $>$5$\sigma$ we find further 8 flares on 8 stars (q1 Eri, HD~4308, HD~21209, HD~42618, HD~65907, HD~157347, HD~199288, and $\psi$ Ser) from which 3 are of spectral type K, 4 are of spectral type G and 3 are of spectral type F. Taking also this definition of a flare into account does not change the result that there is only very little flaring in the data. Calculating a flare incidence yields $\sim$ 1\% for K-stars, $\sim$ 3\% for G-stars, and $\sim$ 10\% for F-stars.\\
In Fig.~\ref{Fig5} we show the variation of the H$\alpha$ pseudo EW as a function of $\log L_{\mathrm{X}}$ of stars of our sample with more than 20 spectra in the Polarbase and HARPS archives. From top to bottom we show the variation of the H$\alpha$ pseudo EW for dF, dG, and dK. The blue filled circles represent stars from the HARPS archive whereas the red filled squares represent stars from the Polarbase archive. We note here that we can only plot stars which have measured X-ray luminosities. The X-ray luminosities are taken from \citet{Hinkel2017, Guedel2007, Gaidos1998, Boller2016}. One can see that all spectral types show an expected behaviour, namely increasing chromospheric variability with increasing X-ray luminosity.

\section{Discussion}
\subsection{Flares}
\label{flarediscuss}
In section~\ref{flaresres} we have presented the only flares which we have found in the data. The total number of flares found on 425 stars in more than 3700~hours of on-source time is 11, if we take both flare definitions into account, which is rather small. The expected number of flares is significantly higher. The power-law by \citet{Audard2000}, from which the expected flare rates are calculated, is based on flare detections by the Extreme UltraViolet Explorer (EUVE). This power-law gives the number of EUV flares ($>$10$^{32}$~erg) per day in dependence of stellar X-ray luminosity. If the stellar X-ray luminosity of a star is known then the flare rate of this star can be calculated. In Table~\ref{appendixtable} we give the flare rates calculated with the power-law from \citet{Audard2000} for flares $>$ 10$^{32}$ erg. Such energies correspond to solar X-class flares. \citet{Poetzi2015} showed a relation of solar flares detected in X-rays by the Geostationary Operational Environmental Satellites (GOES) and H$\alpha$ by the Kanzelh\"ohe Solar Observatory (KSO). According to their Figure~1 a strong correlation (H$\alpha$ flare importance class and GOES X-ray flux), based on 2832 events and ranging from C-X class, exists between both wavelength regimes during flares. 
It is therefore even more surprising that we have detected only very few weak flares in the data. If one would determine the relation shown in \citet{Poetzi2015} for the Sun as a star i.e. integrated over the solar disk then the correlation would break down as the solar H$\alpha$ flare areas are too small relative to the solar disk to result in a significant signal to be seen (see section~\ref{solarfulldisk}). Young main-sequence stars are supposed to have larger active regions and stronger magnetic fields and should therefore also produce stronger H$\alpha$ flares as flares known from the Sun. \\
The distribution of solar/stellar flares follows a power law of the form $\frac{dN}{dE} = kE^{-\alpha}$ \citep[e.g.][and references therein]{Audard2000} with the corresponding cumulative distribution $N(E > E_{\mathrm{min}} = \frac{k}{\alpha-1}E_{\mathrm{min}}^{1-\alpha}$). Here, N is the number of flares exceeding a certain energy E$_{min}$, k is the proportionality factor, and $\alpha$ is the flare power-law index which defines the slope in a flare frequency distribution (FFD). For H$\alpha$ this yields

\begin{equation}
\frac{dN}{dE_{\mathrm{H\alpha}}} = k_{\mathrm{H\alpha}}E_{\mathrm{H\alpha}}^{-\alpha} = \frac{dN}{dE_{\mathrm{XUV}}} \times \frac{dE_{\mathrm{XUV}}}{dE_{\mathrm{H\alpha}}} 
\end{equation}

\noindent To determine the term $dE_{\mathrm{XUV}}/dE_{\mathrm{H\alpha}}$ we use the relation between X-ray and H$\gamma$ luminosities L$_{\mathrm{X}}$=31.6$\times$L$_{\mathrm{H_{\gamma}}}$ and the Balmer decrement F$_{H\alpha}$=3$\times$F$_{H\gamma}$, relating H$\alpha$ and H$\gamma$ fluxes, from \citet{Butler1988}. The luminosity relation can also be written in terms of energy (L$_{X}$=31.6L$_{H\alpha}$ $\rightarrow$ E$_{X}$=31.6E$_{H\alpha}$), assuming a similar flare duration for both XUV and H$\alpha$ wavelength ranges. L$_{X}$ in \citet{Butler1988} corresponds to a spectral region of $\sim$3-300\AA{} whereas E$_{XUV}$ from \citet{Audard2000} refers to a spectral region of $\sim$1-1240\AA{}. Therefore we use the conversion factor 2.6 of EUV(465-794\AA) to SXR(6-280\AA) flux from \citet{Butler1990} to account for the difference. This gives then E$_{3\text{-}794\si{\angstrom}}$=E$_{3\text{-}300\si{\angstrom}}$+2.6E$_{6\text{-}280\si{\angstrom}}$. As E$_{3\text{-}300\si{\angstrom}}$ and E$_{6\text{-}280\si{\angstrom}}$ refer to a similar wavelength range we can write E$_{3\text{-}794\si{\angstrom}}\approx$3.6E$_{3\text{-}300\si{\angstrom}}$. To account for the missing spectral range of $\sim$800-1240\AA{} we introduce an additional conversion factor c. Therefore we can write E$_{XUV}\approx$c$\times$3.6E$_{3\text{-}300\si{\angstrom}}$. The conversion factor c must be $>$1 as E$_{3\text{-}794\si{\angstrom}}$ $<$ E$_{XUV}$. Combining all these relations gives then $dE_{\mathrm{XUV}}/dE_{\mathrm{H_{\alpha}}}$=c$\times$37.9. Inserting this in Eq.~1 gives
\begin{equation}
\frac{dN}{dE_{\mathrm{H\alpha}}} = k_{\mathrm{XUV}} E_{\mathrm{XUV}}^{-\alpha} \times c \times  37.9 = k_{\mathrm{XUV}} \times (c\times37.9)^{1-\alpha} \times E_{\mathrm{H\alpha}}^{-\alpha}
\end{equation}
\noindent and
\begin{equation}
k_{\mathrm{H_{\alpha}}} = k_{\mathrm{XUV}} \times (c\times37.9)^{1-\alpha}
\end{equation}
\noindent and finally
\begin{equation}
\label{finaleq}
\frac{N_{\mathrm{H\alpha}}(E > 10^{32}~\mathrm{erg})}{N_{\mathrm{XUV}} (E > 10^{32}~\mathrm{erg})} = \frac{k_{\mathrm{H\alpha}}}{k_{\mathrm{XUV}}} = (c\times37.9)^{1-\alpha}
\end{equation}
As the conversion factor c is greater than one and the exponent (1-$\alpha$) is lower than one, the determined H$\alpha$ flare rate represents an upper limit. Typical values for $\alpha$ lie in the range of 1.5 .. 2.5 \citep{Audard2000}.\\
\citet{Audard2000} found a power-law relation between $N_{\mathrm{XUV}} (E > 10^{32}~\mathrm{erg})$ and the stellar X-ray luminosity. By combining it with Eq.~\ref{finaleq}, we can estimate the predicted H$\alpha$ flare rates.                                                
\begin{figure}
  \centering
  \includegraphics[width=\columnwidth]{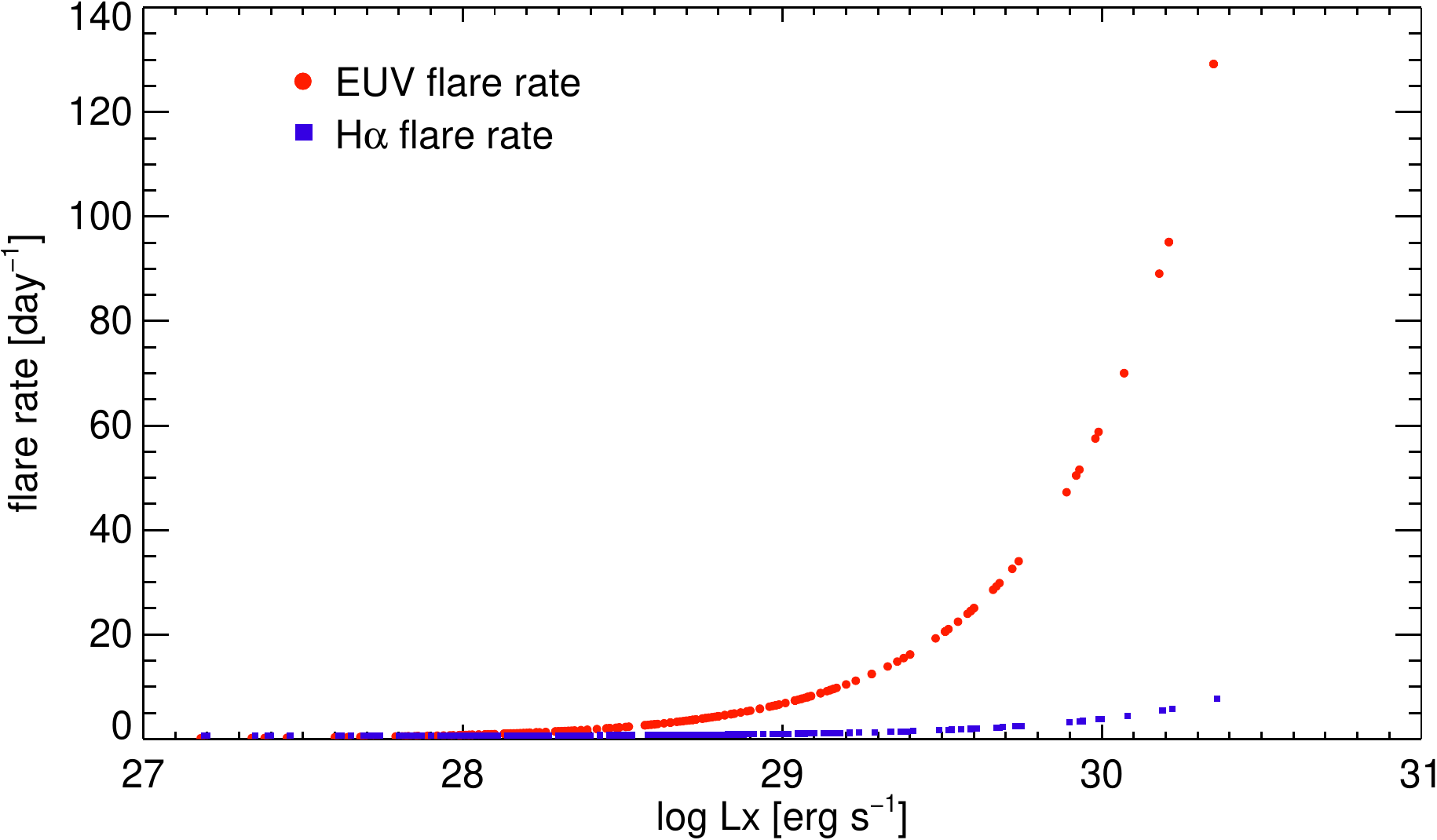}
   \caption{
   Flare rates of the target stars of the present study deduced from the EUV flare power law from \citet{Audard2000} (red dots) and flare rates deduced from the H$\alpha$ power law (blue squares).}
         \label{Fig6}
   \end{figure} 
\noindent \citet{Butler1993} find for a flare on the RS Canum Venaticourm star II~Peg a lower Balmer decrement of H$\alpha$=2$\times$H$\gamma$. As the Balmer decrement of H$\alpha$=3$\times$H$\gamma$ was deduced from a number of solar and stellar flares \citep{Butler1988} we use this Balmer decrement in our calculations. In Fig.~\ref{Fig6} we show the flare rates of the targets stars as deduced from the EUV power law (red dots) and from the adapted H$\alpha$ power law (blue squares). The EUV flare rates range from $<$1 up to $\sim$130 whereas the H$\alpha$ flare rates do not show values beyond 10 per day. However, there are only 8 stars which have a predicted H$\alpha$ flare rate of $>$ 1 flare per onsource time and none of them showed any flare. The predicted H$\alpha$ flare rates of the stars showing flares are below 1 flare per onsource time. However, the predicted flare rates are valid for flare energies $>$10$^{32}$~erg. The flare energies of HD~189733 and $\iota$~Hor are below 10$^{32}$~erg (see section~\ref{flaresres}). As the other observed flares consist of only one data point (see section~\ref{flaresres}) we do not determine their energies, as this determination is not reliable and useful.\\
With the prominent exoplanet search missions CoRoT \citep{Baglin2006} and Kepler \citep{Borucki2010}, both operating in the optical, it became possible to deduce flare rates from the monitoring data obtained by those missions. Superflare detections have been presented for solar-like \citep{Maehara2012, Notsu2013, Shibayama2013} and also late-type stars \citep{Maehara2014, Candelaresi2014}. Moreover, both long- and short-cadence Kepler data, were used to determine flare rates of late-type main-sequence stars \citep{Balona2015, Davenport2016, VanDoorsselaere2017, Yang2019}. \citet{Yang2019} present flare incidences which is the fraction of investigated stars which show flares. They find $<$1\% for F stars, $\sim$1.5\% for G stars, and $\sim$3\% for K stars. Flare incidences from \citet{Balona2015} and \citet{VanDoorsselaere2017} are a factor $\sim$1.5-5 higher than the flare incidences presented in \citet{Yang2019}, which the latter explained by the identification of false-positives . However if we compare our estimated flare incidences from section~\ref{flaresres} to the flare incidences from \citet{Yang2019} the values for G- and K-stars differ by a factor of $\sim$3. The comparison of our F-star flare incidence ($\sim$10\%) is of low significance because our F-star sample is small, especially compared to our G- and K-star sample. From that comparison we are in between the values represented in \citet{Yang2019, Balona2015, VanDoorsselaere2017}. But we have to keep in mind that our study derived flares from a pseudo H$\alpha$ EW with a fixed wavelength window of 2\AA{} and Kepler flares are derived from photometry with a passband of nearly 5000\AA{}. In section~\ref{cmediscuss} we discuss several issues related to the non detection of CMEs, however some of them (target stars, data) concern also the very low flare rate estimated in this study.\\ 
\citet{Butler1988} and \citet{Butler1993} showed that for simultaneously observed X-ray and H$\gamma$ flares. To get an idea what number of flares with corresponding H$\alpha$ fluxes we should have detected we adapt the power-law by \citet{Audard2000} to the H$\alpha$ domain. For this purpose we use the above mentioned relations between stellar flare X-ray and H$\gamma$ flux from \citet{Butler1988}, \citet{Butler1993}, and \citet{Haisch1989}.

\subsubsection{Solar disk-integrated H$\alpha$ flares}
\label{solarfulldisk}
As we detected only very few and less energetic flares (see section~\ref{flaresres}) we pose the question if one could detect a solar flare in H$\alpha$ if the Sun is seen as a star, i.e. in disk-integrated light. As our target sample is mainly  comprised of solar-like stars this is reasonable. Therefore we searched the archives of the Big Bear Solar Observatory (BBSO) and the Kanzelh\"ohe Solar Observatory (KSO) for solar H$\alpha$ full disk images of the most energetic solar flares (down to X4.0, selected from the GOES SXR flare list\footnote{\url{https://hesperia.gsfc.nasa.gov/goes/goes_event_listings/}}). We had no luck with the search in the KSO archive as we have found only 3 flares from our list and the data of those 3 flares were unfortunately contaminated by clouds. 
\begin{figure}
  \centering
  \includegraphics[width=\columnwidth]{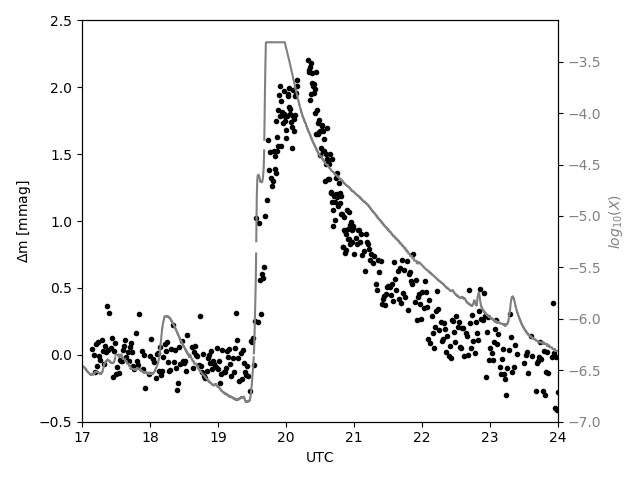}
   \caption{Light curve of the Sun in H$\alpha$ covering a flare on the 4$^{th}$ of November 2003. 
   The black dots denote the light curve from BBSO data. The grey solid line denotes the X-ray light curve from the GOES satellite.}
         \label{solarfig}
   \end{figure} 
In the BBSO archive we found only 5 flares from our list. We focus here on the highly energetic solar X28.0 flare from 4$^{th}$ of November 2003 peaking at 19:53~UT which has been captured by BBSO. Data at BBSO was taken from 17:18~UT until 00:26~UT, nearly covering the complete X-ray flare, with about a 1~minute cadence. 
This flare was located at the solar limb. In principle this can be measured by integrating the solar disk flux and subtracting the scattered background light. However, when 
\begin{figure}
  \centering
  \includegraphics[width=\columnwidth]{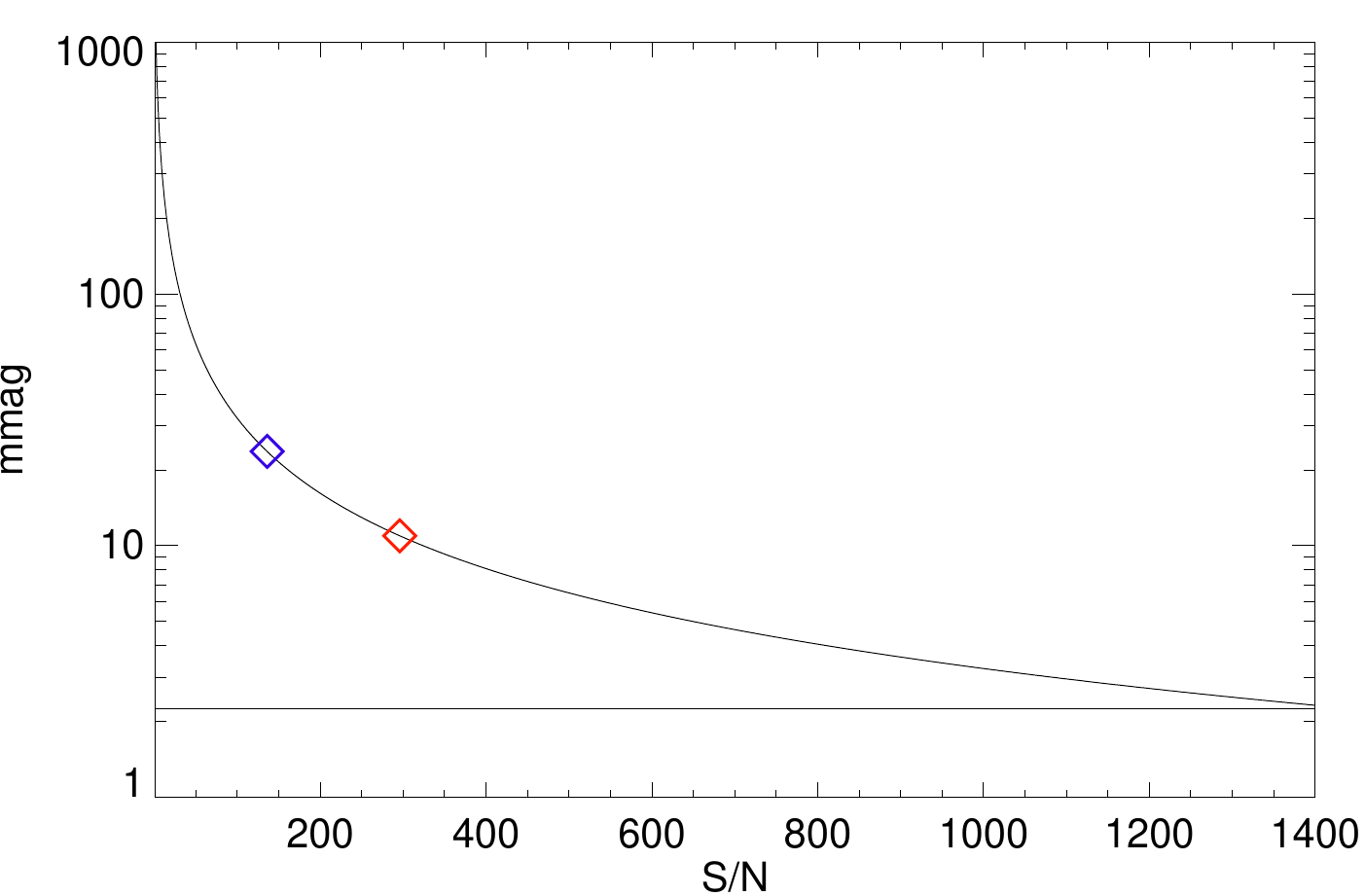}
   \caption{
   S/N of data versus mmag of an enhancement caused by flares in the H$\alpha$ line core. The blue and red diamond symbols denote the mean S/N of the HARPS and Polarbase data of the present study, respectively. The horizontal line at 2.25~mmag denotes the enhancement in the 
H$\alpha$ linecore due to the most energetic flare in the last solar cycles, which is the Halloween event on 4th of November 2003.}
         \label{Fig6a}
   \end{figure} 
doing this, no flare was visible in the light curve. 
To integrate just the flare area, we first need to determine the center of the solar disk in each image. This is done by fitting the disk at a fixed count level above the off disk background. The center is then refined using the position of a small sunspot, taking into account its positional shift due to solar rotation. We then integrated a square area around the flare, subtracting the local off solar background. This gave a flat flux distribution before the flare and a steep rise at the flare time. The average flux in the flat part of the light curve was subtracted from the flare area integration, leaving just the additional flux coming from the flare. We then determined the average solar full disk flux before the flare and added this to the flare flux (see Fig.~\ref{solarfig}). 
The rise of the light curve at the start of the flare for both datasets can clearly be seen. In the jpeg data we can also see the exponential drop off that well follows the X-ray light curve (grey solid line). The enhancement of the solar H$\alpha$ flux due to the strongest solar flare recorded so far amounted to $\sim$2.25~mmag. For the H$\alpha$ observations BBSO uses a Zeiss Lyot filter with a bandwidth of 0.25\AA{}. The pseudo EW deduced in the present study were determined from a 2\AA{} wide window. We expect that the solar flare enhancement would be lowered if a filter with a larger bandwidth would have been used. Therefore the determined solar flare enhancement is treated as an upper limit when compared to the stellar spectroscopic observations of the present study. 
In Fig.~\ref{Fig6a} we show the detectability of flare enhancements (in units of mmag) in dependence of S/N of the stellar spectroscopic data. One can see that we could not have detected a flare which causes a 2.25~mmag enhancement in the Polarbase and HARPS data. The main reason for the weak enhancement detected for the solar X28 flare is that the area of the H$\alpha$ brightening is very small with respect to the solar disk. 

\subsection{Expected CME rates}
\label{expcme}
To compare with the observations, we apply a semi-empirical model to estimate CME rates observable in H$\alpha$. The modeling approach is described in detail in \citet{Odert2019}, here we briefly 
summarize the adopted method. First, we estimate the expected CME rates of the stars from their X-ray luminosities using the empirical model described in \citet{Odert2017}. This empirical model combines a relation between X-ray luminosity and stellar flare rates from \citet{Audard2000} and relations between flare energies and CME masses from the Sun \citep{Drake2013}. It takes into account the solar flare-CME association rate in dependence 
of flare energy \citep{Yashiro2006}. This approach assumes that the relations between flares and CMEs on the Sun and other solar-like stars are similar, which is likely a reasonable assumption for Sun-like 
stars of low to moderate activity levels, which comprise the majority of our sample. The empirical model provides an extrapolated CME rate (and CME  mass distribution) for any star with given X-ray luminosity.
In \citet{Odert2019} we then use a simplified solution to the radiative transfer equation \citep[e.g.][]{Heinzel2015} to estimate the H$\alpha$ fluxes for the range of CME masses estimated for every star. We consider both 
emission (i.e., prominence) and absorption (i.e., filament) geometries when calculating the disk-integrated fluxes. As the H$\alpha$ flux is emitted by the neutral material of the prominence embedded in the 
CME core, we assume for simplicity that the total CME mass is composed of neutral hydrogen. With this assumption we obtain the maximum expected H$\alpha$ signatures, which overestimates the extrapolated number 
of observable CMEs. The detectability of the achieved signals is estimated from the average S/N of the spectra for each star, such that the peak H$\alpha$ fluxes of the CMEs have to exceed the noise of the spectra.
This potentially observable number of CMEs is further reduced by projection effects. For instance, we are only able to observe the line-of-sight velocities of CMEs in spectra. To account for the reduction of 
observable CMEs due to velocity projection, we use the CME mass-velocity relation from the Sun \citep{Drake2013} and calculate the fraction of CMEs with sufficient line-of-sight velocity components to be detected in the 
spectra using a Monte Carlo method. Hereby, we assume that CMEs are randomly ejected radially in all directions to obtain the fraction of CMEs with sufficiently high line-of-sight velocities (${>}100$\,km\,s$^{-1}$).
Furthermore, we take into account the limited duration of absorption signals, which may be too short to be detected if a CME moves off the disk faster than the typical exposure time of the spectra. This is also included in 
the Monte Carlo approach by calculating the duration in front of the stellar disk depending on the ejection trajectory and CME velocity. Application of this semi-empirical model yields then the maximum possible CME rates observable 
in H$\alpha$ for all target stars with known X-ray luminosity. We compare these extrapolated CME rates with the upper limits from observations in the next section.

\subsection{CMEs}
\label{cmediscuss}
We searched the Polarbase and HARPS phase 3 archive for spectra of dF-dK stars extracted from the input list described in section~\ref{targetstars}. The final target list contains 456 stars from which 31 stars have both, HARPS and Polarbase data. From those 425 stars we found no signature of stellar CMEs in form of blue- or red-wing asymmetries in more than 3700~hours of pure on-source time. Reasons for non detections are manifold and include also biases connected to target selection and available data.\\
\textbf{\textit{Targets:}} Our target sample is mainly based on the F-K star list presented in \citet{Hinkel2017} with additions from \citet{Gaidos1998} and \citet{Guedel2007}.
The input list from \citet{Hinkel2017} includes F-K main-sequence stars within a distance of 30~pc. Although their list has 80$\%$ completeness it represents a profound base to work with. The first bias is introduced by the target sample. The stars differ in age or $\log L_{\mathrm{X}}$. Therefore older or less X-ray luminous stars should not show many CMEs in contrast to the younger stars. In the target sample we find 33 stars between $\log L_{\mathrm{X}}$ 27-28~erg~s$^{-1}$, 96 between $\log L_{\mathrm{X}}$ 28-29~erg~s$^{-1}$, 51 between 29-30~erg~s$^{-1}$, and 8 between 30-31~erg~s$^{-1}$. For the remaining stars we have no information on X-ray luminosity. Another bias connected to the target sample is that we have different spectral types in the sample. The continuum level typically changes from a dF- to a dK-type star at H$\alpha$, for a dF star the continuum level around H$\alpha$ is higher than that for a dK star.\\
\textbf{\textit{Data:}} This connects directly to the next source of biases, namely the data. Assuming that an erupting stellar filament causes a certain flux, the detection of this flux on a dF star needs higher S/N in the data than for a dK star \citep[see Fig.~1 in ][]{Odert2019}. The spectra of the stars yield various S/N values. On average the dF and dG stars of the sample yield similar values of $\sim$~200 and the dK stars yield on average a S/N of $\sim$~80. These differences in S/N are considered when calculating the maximum expected observable CME rates. Therefore there is quite a number of stars in Table~\ref{appendixtable} which yield no expected observable CMEs during their observing time although the stars have a measured $\log L_{\mathrm{X}}$. However, if we use Fig.~1 in \citet{Odert2019} to check which CMEs we could have detected with the S/N of the data used in the present study, then we see that we could have detected CMEs with masses above 5$\times$10$^{16}$~g for dF stars. For dG stars we could have detected CMEs with masses above 3$\times$10$^{16}$~g and for dK stars we could have detected CMEs with masses above 4$\times$10$^{16}$~g.\\
The S/N of the spectra is a main limiting factor in the detection of stellar CMEs. The data we use in the present study originate from Echelle spectrographs with very high resolving powers leading to high spectral resolution of $\Delta\lambda\sim$ 0.05 (HARPS) and 0.08~\AA{} (ESPaDOnS/Narval). To see if we find signatures of stellar CMEs in higher S/N data we bin the spectra by a factor of 10 in wavelength, thereby reducing the spectral resolution down to $\Delta\lambda\sim$ 0.5 (HARPS) and 0.8~\AA{} (ESPaDOnS/Narval). This enhances the S/N of the data by a factor of $\sim$3. Re-analysing the spectra even rebinned by a factor of 10 did not show signatures of stellar CMEs. As the rebinned spectra have a S/N of a factor $\sim$3 higher than before also the corresponding CME mass ranges (see paragraph above) are lowered by a factor of $\sim$3.\\
The most massive solar CMEs have masses of $\sim$10$^{17}$g \citep{Yashiro2009} and the so far detected stellar CMEs have masses (using the method of Doppler shifted emission/absorption) in the range of 10$^{15}$-10$^{19}$~g \citep{Gunn1994, Guenther1997}. Using the method of X-ray absorptions stellar CME masses up to 10$^{23}$~g are reported \citep[cf.][and references therein]{Moschou2019}, but the majority of those stars are no late-type main-sequence stars.  The deduced minimum detectable CME masses from the S/N values are in the ranges of the stellar CMEs detected by the method of Doppler shifted emission/absorption. Therefore it could have been possible to detect moderate to high mass stellar CMEs with the Polarbase and HARPS spectra.\\
The second bias related to the data is introduced by the total on-source. Each data set has a different on-source time and a different number of consecutive spectra (see Fig.~\ref{Fig2}). In the target star sample there are 48 stars which have an on-source time $>$ 1~day and 39 stars with an on-source time between 0.5 .. 1~day. The remaining stars have on-source times $<$ 0.5~ days. \\
\begin{figure*}
  \centering
  \includegraphics[width=15cm]{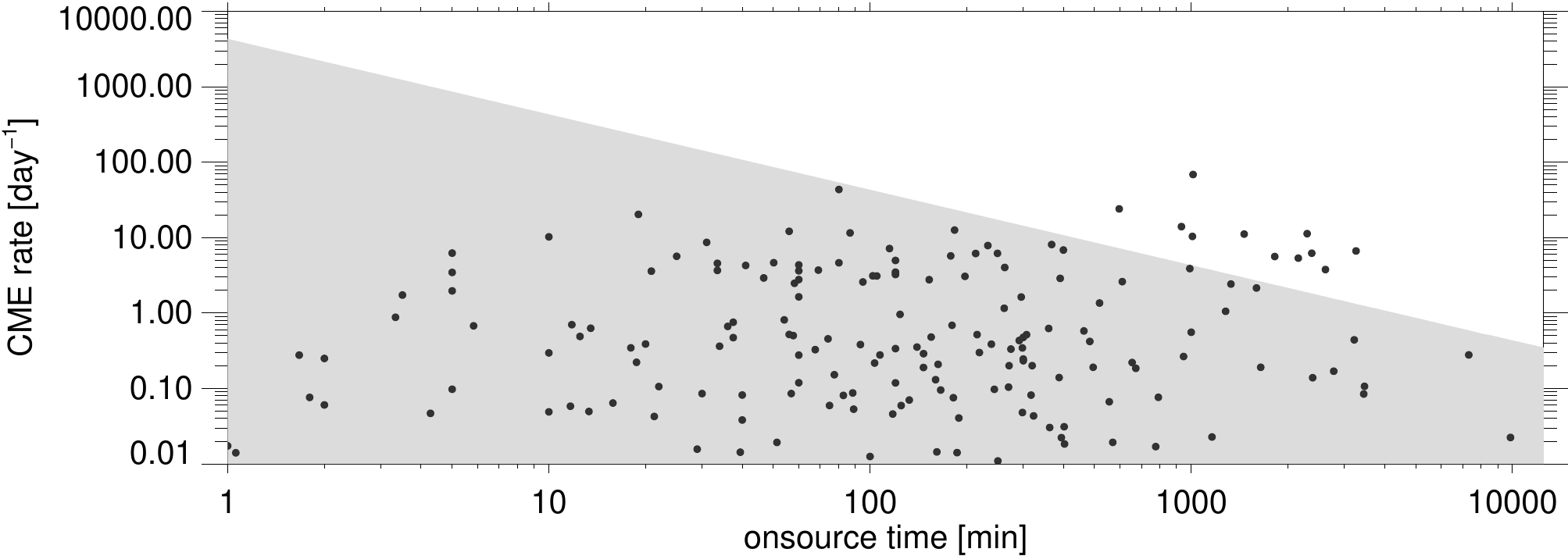}
\caption{On-source time versus CME rate. The grey shaded area denotes the upper limits of the non detections of stellar CMEs in the Polarbase and HARPS data. The black dots denote the maximum expected observable CME rates calculated using the semi-empirical model of \citet{Odert2019}.}
\label{Fig7}
\end{figure*} 
As we have detected no signatures of CMEs in the spectra the observed CME rates are upper limits only (see section~\ref{CMEres}). Stars with a short on-source time yield consequently large upper limits and stars with long on-source times yield small upper limits, as the upper limit of the observed CME rate is a function of on-source time only. The upper limits can be used to constrain the modelled CME rates. This is shown in Fig.~\ref{Fig7} where we plot the on-source time versus the observed upper limits. This is marked by the grey shaded area. As expected and described above the upper limits decrease with increasing on-source time. We also overplot the maximum expected modelled CME rates (black dots). With this plot we are able to constrain the maximum expected modelled CME rates, but only for those stars where the maximum expected modelled CME rates exceed the upper limits derived from observations. We find \textbf{11} stars which have up to 1.5 order of magnitude higher expected CME rates than the upper limits derived from observations (stars with an expected CME rate $>$ 2~day$^{-1}$ and an on-source time $>$ 500~min). Those stars have higher log~Lx (in the range of 29-30~erg~s$^{-1}$, see also Table~\ref{appendixtable}). The S/N of the observations of those stars are also higher than on average (cf. Fig.~\ref{Fig2} and Table~\ref{appendixtable}).\\
However, for the majority of stars, the upper limits exceed the expected CME rates. We want to note here that the majority of stars has on-source times $<1$~day and therefore the upper limits show large values, therefore the expected CME rates lie below the upper limits. With increasing on-source times the observed upper limits approach the modelled CME rates.\\
Apart from biases in target selection and data we discuss also other reasons for non-detections of stellar CMEs.\\
\textbf{\textit{Doppler dimming:}} \citet{Hyder1970} reported on brightness variations of moving prominences. This is described as so-called ``Doppler dimming/brightening'' which means that in the frame of the moving prominence the solar radiation appears Doppler shifted and therefore moving prominences appear brightened or dimmed depending on the incident stellar spectrum as their emission is dominated by scattering. In a simplified two-level atom, Doppler dimming occurs when the incident spectral line is in emission and Doppler brightening occurs when the incident spectral line is in absorption \citep{Hyder1970, Labrosse2010}. According to \citet{Heinzel1987} and \citet{Gontikakis1997} a more realistic scenario is obtained when using a multi-level atom model in which the energy level populations depend on various line transitions from the Lyman, Balmer etc. series. Taking this into account leads to a more complex behaviour with increasing velocity that can lead to either a brightening and/or dimming.\\
According to the simplified approach of a two-level atom the signature of a moving structure would get brightened around all but the youngest solar-like stars. According to the more realistic approach of a multi-level atom, the signature of a moving structure would get brightened up to a certain velocity. If the moving structure is faster than that velocity the opposite effect occurs and it would get dimmed \citep[cf. Fig.~5][]{Heinzel1987}. The modelling in \citet{Heinzel1987} was applied to solar moving prominences and our target sample is mainly comprised of solar-like stars, showing at least the Balmer lines in absorption. We would therefore expect that Doppler brightening decreases for moving structures, for stars showing H$\alpha$ in absoprtion as the Sun, with larger velocities.
For the Sun, Doppler brightening may enhance the intensity up to factors of 2-3 for optically thin prominences and a bit less for optically thick prominences \citep{Heinzel1987}.
To evaluate the influence of Doppler dimming on our target stars, dedicated modelling would be required, which is beyond the scope of the current study, but would be highly desirable. However, the semi-empirical model of \citet{Odert2019} includes a simple estimate of this effect in the the calculations of the CME fluxes. Thus, this effect could have contributed to the non-detection, as most extrapolated maximum observable CME rates from this semi-empirical model are below the observational upper limits (see Fig.~\ref{Fig7}).\\
\textbf{\textit{Projection effects and inclination}}: CMEs are ejected in random directions. Therefore one can measure projected velocities only and it is not possible to classify slow events as stellar CMEs. However, in the calculation of the maximum expected CME rates projection effects are considered cf. section~\ref{expcme}.\\
The Sun has spots occurring around the equator. CMEs are ejected, or have their source locations, close to the equatorial plane during solar minimum \citep{Wang2011}. During solar maximum CMEs occur also at higher latitudes on the Sun \citep{Harrison2018}. According to \citet{Harrison2018} there is evidence that CMEs also occur aside from active regions. \citet{Gopalswamy2010} show source locations of CMEs in the period 1996-2008. From their Fig.~9 one can see that, in compliance with \citet{Harrison2018}, there are source locations reaching high latitudes, but the majority of source locations occurs at low- and mid-latitudes. However, young and active stars, which typically have short rotation periods, show a trend towards dominant polar spots and less dominant mid-latitude spots \citep{Strassmeier2009}. If one observes a star which has polar and mid-latitude spots and an inclination of i=90$^{\circ}$ then one would possibly not see CME signatures because there would be no or only a very small line-of-sight velocity component. For the majority of stars the inclination of their rotation axes remains unknown. Assuming that stellar rotation axes of late-type main-sequence stars are oriented randomly, this should affect the search for CMEs on several hundreds of stars only little.\\
\textbf{\textit{Ionization:}} Another issue related to filament detectability is the ionization of the filament material. \citet{Howard2015} measured how long one can see an eruptive prominence on the Sun in H$\alpha$ and they found that at a distance of $\sim$7R$_{\odot}$ the dense filament material becomes completely ionized or at least the Thomson scattering process becomes dominant. It is known that coronae of young and active stars are hotter and the X-ray luminosity is much higher than on the present-day Sun \citep[see e.g.][]{Ribas2005, Guedel2007}. Thus, the ionization of filament material could be more effective than on the Sun. On the other hand it has been shown that stellar filaments located at several stellar radii from the star are still visible in H$\alpha$ \citep[][and several others]{CollierCameron1989a}, but it remains unknown if the stronger ionization around those stars (higher X-ray and EUV environment) prevent detection of their eruption in Balmer lines. \\
\textbf{\textit{Magnetic field}}: The magnetic field of a star plays a crucial role in the generation and ejection of CMEs from a star. The reconnection of magnetic field lines yields the energy needed to eject millions of tons into the astrosphere of a star. On the Sun two-ribbon flares and CMEs are closely connected, both get their energy from reconnecting magnetic field lines. Therefore a close correlation between both phenomena exists, which reaches 100$\%$ \citep{Yashiro2009} for so-called X-class flares which belong to the most energetic class of flares on the Sun. In rare cases this relation does not hold and the magnetic field itself prevents a CME to be ejected. One of such cases happened in 2014 on the Sun, where an unusually large active region had formed. In total 6 X-class flares were produced by that region but without any CMEs \citep{Sun2015, Thalmann2015}. It has been suggested that a stronger overlying magnetic field together with weaker non-potentiality leaded to confinement and therefore no CMEs. From a statistical point of view such events are rare on the Sun. \citet{Drake2016} suggested that strong overlying magnetic fields, similar as in the solar case described above, might be responsible for the so far sparse detection of CMEs on stars. This is a reasonable suggestion as young and active stars are known to exhibit much stronger photospheric magnetic fields (up to several kG) as the Sun (1-2G). \citet{Alvorado2018} found that an overlying large-scale magnetic field of 75G is sufficient to confine strong CMEs on the Sun or eruptive flares up to a GOES X-ray flare class of X20. When trying to extrapolate these findings to active stars the stellar characteristics drastically change. If we recall that a CME core is in many cases a filament, at least on the Sun, then we first need to compare solar and stellar filaments. On the Sun filaments have heights of ~20Mm ($\sim$0.03~R$_{\odot}$) on average, which means that they are located rather closely to the solar surface. On stars prominences have been found with heights of a few stellar radii \citep[see e.g.][and references therein]{Villarreal2019}. Even if young and active stars exhibit strong photospheric magnetic fields the question is if those fields are still strong enough to confine a filament eruption at such heights. The magnetic field strength from a star declines with ($R_{\star}$/r)$^{n}$, with r being the distance from the star in units of stellar radii and e.g. n=3 for a dipolar field. If we calculate the magnetic field strength for an active star with a dipole field at the height of a stellar prominence of 2R$_{\star}$ assuming a photospheric magnetic field strength of 2kG then at 2R$_{\star}$ the magnetic field strength has reduced to 250G. Only for very few stars magnetic fields have been reconstructed  \citep[see e.g.][]{Reiners2012}, those can be quadrupole or even more complex magnetic fields \citep[see e.g.][]{Donati2009}. For a quadrupole field the magnetic field strength would have reduced to 125G and for an octupole field to 63G at 2R$_{\star}$. Moreover, one should also keep in mind that on active stars not only the overlying magnetic field is stronger but also the sizes of stellar spots and their magnetic field strengths are larger which lead then to the observed larger flare energies \citep[e.g.][]{Audard2000}. The available magnetic energies could still lead to eruptions even in stronger magnetic fields.\\
But even if the overlying magnetic field prevents stellar CMEs from eruption one should at least see in optical spectra the initial rise phase until the filament plasma is prevented by the magnetic field from eruption. But in the present study we did not even find extra emissions/absorptions with small velocities which may represent such a scenario.





\section{Conclusions}

The goal of the present study was to determine the frequency of CMEs of solar-like stars. In more than 3700 hours of on-source time of 425 stars of spectral type dF-dK we found no signature of stellar CMEs in optical spectra. Aside from the biases introduced by the target selection and data, where we identified on-source time and S/N to be the most important ones, we discussed several issues (Doppler dimming, projection effects and inclination, ionization) which could hinder the detection of stellar CMEs. The magnetic field of the individual stars plays also a crucial role as a strong overlying field can prevent filaments from eruption.\\
For most of the target stars the observed upper limits of the CME rates are larger than the maximum expected CME rates. For a rather small fraction of the target sample ($\sim$4\%) we found that the maximum expected CME rates exceed the observed upper limits.\\
We found negligible flare activity of the stars in the target sample. According to X-ray/EUV observations of young solar-like stars \citep{Audard2000} we know that flaring happens more often on those stars. We conclude therefore, apart from the biases discussed above, that the flare brightened H$\alpha$ regions are too small relative to the stellar disk to be recognized in stellar spectral H$\alpha$ time series.  \\
According to the S/N of the data we could have been able to detect CMEs with masses in the upper range of the solar CME mass distribution (10$^{16}$ - 10$^{17}$g) or above. All CMEs with masses below that remain invisible for the present study. For investigating stellar CMEs with lower masses ($<$10$^{16}$) data with higher (in the case of dF stars much higher) S/N would be needed. Such data quality is available only for very few spectra in the archives. Given that our observational capabilities suffice, our results indicate that CMEs in the upper solar range of masses are not very frequent.\\
Although archival data represent a rich data pool for the determination of sporadic activity phenomena such as flares and CMEs, they offer mainly short observational time series. This of  course introduces biases. Another approach involving archival data is the usage of spectroscopy surveys such as e.g. the Sloan Digital Sky Survey (SDSS). Although the biases related to archival data remain but the number of stars to be investigated is much higher than the number of stars investigated in the present study. An alternative way to characterize stars in terms of flares/CMEs is multi-object spectroscopy. This is not a new concept as several authors have already used multi-object spectroscopic data to search for sporadic activity phenomena \citep[e.g.][]{Guenther1997, Leitzinger2014, Korhonen2017}. This observational approach can be further optimized by using telescopes/instruments ensuring a large field of view ($>$1 deg$^{2}$) together with a high number of young pre/main-sequence stars, usually located in open clusters or star forming regions.\\
The main problem when characterizing stellar activity phenomena is that a large amount of data i.e. a lot of observing time at telescopes is required. Using the method of Doppler-shifted emission/absorption as a signature of CMEs requires spectroscopic observations. Obtaining observing time on telescopes for several weeks or even longer is unrealistic as this would block a telescope completely for one science case. Therefore an activity alert system is a promising concept. Such a concept has been presented by \citet{Hanslmeier2017}. The system requires a rather small telescope which monitors selected regions in the sky, containing a high number of late-type main-sequence stars, photometrically. If a flare is detected an alert is sent in real time to collaborating observatories performing spectroscopic follow-up observations. This system represents an efficient way to characterize stellar activity phenomena, especially stellar CMEs. 

\section*{Acknowledgements}

M.L., P.O., R.G., and F.K. acknowledge the Austrian Science Fund (FWF): P30949-N36 for supporting this project. M.L. and P.O. acknowledge the Austrian Space Applications Programme of the Austrian Research Promotion Agency FFG (ASAP-14 865972, BMVIT). M.L., P.O., R.G., K.V., L.K., A.H., and Zs.K. acknowledge the financial support of the Austrian-Hungarian Action Foundation (95\"ou3, 98\"ou5, 101\"ou13). A.H. acknowledges the Austrian Science Fund (FWF): I3955 for supporting this project. K.V. is supported by the Bolyai J\'anos Research Scholarship of the Hungarian Academy of Sciences. K.V., L.K., and Z.K. are supported by the NKFIH grant K-131508 from the Hungarian National Research, Development and Innovation Office. This work has made use of data from the European Space Agency (ESA) mission {\it Gaia} (\url{https://www.cosmos.esa.int/gaia}), processed by the {\it Gaia} Data Processing and Analysis Consortium (DPAC, \url{https://www.cosmos.esa.int/web/gaia/dpac/consortium}). Funding for the DPAC has been provided by national institutions, in particular the institutions participating in the {\it Gaia} Multilateral Agreement.)




\bibliographystyle{mnras}
\bibliography{Mybibfile} 




\appendix

\onecolumn
\scriptsize
\begin{landscape}
\begin{longtable}{cccccccccccc}
\caption{All target stars from the present study. Spectral types and $\log L_{\mathrm{X}}$ values, are taken from \citet{Hinkel2017, Guedel2007, Gaidos1998, Boller2016}. Distances are taken from GAIA DR2 \citep{Gaiacollaboration2018b}. The star and asterisk symbols denote the origin of the data which is the Polarbase archive ($^\star$) and the ESO phase 3 HARPS archive ($^\ast$). S/N values represent mean values of the corresponding spectral data of each star. The EUV flare rates are calculated after \citet{Audard2000}, the H$\alpha$ flare rates are calulated as described in section~\ref{flarediscuss}, the observed upper limits of the CME rates are calculated as described in section~\ref{CMEres}, and the maximum expected CME rates are calculated after \citet{Odert2019} using an $\alpha$ of 1.8, which is a typical flare power-law index for the solar flare distribution \citep[see e.g.][]{Hudson1991}.. The last column gives the standard deviations of the pseudo EWs of the spectral time series of each star in \%.}\\
TYC & name & spec type &   $\log L_{\mathrm{X}}$   & dist & on-source &  S/N &  flare rate    &     flare rate       & CME rate obs. & max exp. CME rate  &  $\sigma$ from mean\\
    &      &           &                &      &          &     &                &                      &  upper limit  &                    &                \\
    &      &           & [erg~s$^{-1}$] & [pc] &   [min]  &     & EUV [d$^{-1}$] & H$\alpha$ [d$^{-1}$] &  [d$^{-1}$]   &  [d$^{-1}$]        &      [\%]\\
\hline

\endfirsthead
\multicolumn{12}{c}%
{{\bfseries \tablename\ \thetable{} -- continued from previous page}} \\
\multicolumn{1}{c}{TYC} & \multicolumn{1}{c}{name} & \multicolumn{1}{c}{spec type} & \multicolumn{1}{c}{$\log L_{\mathrm{X}}$} & \multicolumn{1}{c}{dist} & \multicolumn{1}{c}{on-source time} & \multicolumn{1}{c}{S/N} & \multicolumn{1}{c}{flare rate} & \multicolumn{1}{c}{flare rate} & \multicolumn{1}{c}{CME rate obs.} & \multicolumn{1}{c}{max. exp. CME rate}  & \multicolumn{1}{c}{$\sigma$ from mean}\\ 
   &      &           &                &      &          &     &                &                      &  upper limit   &                    &     \\
   &      &           & [erg~s$^{-1}$] & [pc] &   [min]  &     & EUV [d$^{-1}$] & H$\alpha$ [d$^{-1}$] &  [d$^{-1}$]    &  [d$^{-1}$]        &    [\%]\\
\hline
\endhead
\multicolumn{11}{|r|}{{}} \\ 
\endlastfoot

TYC1030-4610-1$^{\ast}$ & HD 174080 & K4 & 28.75 & 16.9 & 18.00 & 91.68 & 3.9 & 0.21 & 239.66 & 0.30 & -\\
TYC1049-315-1$^{\ast}$ & HD 180263 & K2IV-V & 27.85 & 26.3 & 48.00 & 57.63 & 0.54 & 0.03 & 89.87 & 0.00 & -\\
TYC107-1139-1$^{\star}$ & HD 3363 & G0~V     & - & 29.7 & 120.00 & 212 & - & - & 35.95 & - & 0.18\\
TYC11-343-1$^{\ast}$ & HD 4256 & K3IV-V & - & 22.4 & 46.67 & 92.63 & - & - & 92.43 & - & -\\
TYC1133-1258-1$^{\ast}$ & HN Peg & G0V & 29.12 & 18.1 & 69.04 & 109.3 & 8.77 & 0.48 & 62.48 & 3.70 & 0.87\\
TYC1133-1258-1$^{\star}$ & HN Peg & G0~V     & 29.12 & 18.1 & 2155.00 & 263.8 & 8.77 & 0.48 & 2 & 5.30 & 1.37\\
TYC1167-1661-1$^{\star}$ & HD 218687 & G0~V(k)  & 29.38 & 25.1 & 120.00 & 274 & 15.48 & 0.84 & 35.95 & 3.50 & 1.16\\
TYC1200-235-1$^{\star}$ & HD 8262 & G2~V     & - & 26.1 & 90.00 & 96.67 & - & - & 47.93 & - & 0.5\\
\end{longtable}

\end{landscape}


\bsp	
\label{lastpage}
\end{document}